\documentclass[12pt]{article}
\usepackage{amsmath}
\usepackage{bezier}
\usepackage{epsf}
\usepackage{epsfig}
\newcommand{\beq}{\begin{eqnarray}}
\newcommand{\eeq}{\end{eqnarray}}
\newcommand{\bea}{\begin{array}}
\newcommand{\eea}{\end{array}}
\def\ba{\begin{eqnarray}}
\def\ea{\end{eqnarray}}
\def\nl{\nonumber\\}
\def\dz{\lambda_Z}
\def\theequation{\arabic{section}.\arabic{equation}}

\newcommand{\litwo}{\mbox{Li}_2}
\newcommand{\lsim}{\raisebox{-0.13cm}{~\shortstack{$<$ \\[-0.07cm] $\sim$}}~}
\newcommand{\gsim}{\raisebox{-0.13cm}{~\shortstack{$>$ \\[-0.07cm] $\sim$}}~}
\begin{document}
\thispagestyle{empty}
 \begin{flushleft}
 DESY 99--165 
 \\
 UG--FT--112/00
 \\
 LC-TH-2000-007
 \\
 hep-ph/0001273
 \\
% December 1999 
% \\
 Version of 28.09.00 
\end{flushleft}
%-----------------------------------
 
 \noindent
 \vspace*{0.50cm}
\begin{center}
 \vspace*{1.5cm} 
{\huge  
Predictions for $Z \to \mu \tau$ 
\vspace*{5mm}
\\
and Related Reactions~$^{\dag}$
%\vspace*{3mm}
%\\%
%
%~$^{a,b}$
 \vspace*{2.cm}               %preprint 
%for LC contribution \vspace*{1.cm}             %for LC contribution
}
%\end{Large}
%-----------------------------------
%\nn
\\
{\large 
J.I. Illana~$^{\ddag}$, M. Jack, and T. Riemann
}
\vspace*{0.5cm}
 
\begin{normalsize}
{\it
Deutsches Elektronen-Synchrotron DESY
\\ 
Platanenallee 6, D-15738 Zeuthen, Germany
}
\end{normalsize}
\end{center}
 
 \vspace*{1.5cm} %preprint 
 \vfill 
%for LC contribution \vspace*{1.cm}            %for LC contribution

\begin{abstract}
We discuss predictions for the lepton-flavour changing decays $Z \to
e\mu, \mu\tau, e\tau$ 
which may be searched for at the Giga--$Z$ option of the Tesla Linear
Collider project with $Z$ resonance production rates as high as 10$^{9}$.
We try to be as model-independent as possible and consider both the
Dirac and Majorana mass cases.
The Standard Model, if being minimally extended by the inclusion of 
light neutrino masses with
some mixings as observed in neutrino oscillation search experiments, predicts
completely negligible rates.
With a more general neutrino content
%In more general schemes 
quite interesting 
expectations may be derived.
\end{abstract}  

%------
 \normalsize
 \vfill
 \vspace*{.5cm}
  
 \bigskip
 \vfill
 \footnoterule
 \noindent

 {\small
$^{\dag}$~Contribution to:  R. Heuer, F. Richard, P. Zerwas (eds.),
 ``Proc. of the 2nd Joint ECFA/DESY Workshop -- Physics Studies for a
Future Linear Collider'',  to appear as DESY report 123F.}
%\ Contribution to the Proc. of the 2nd Joint ECFA/DESY Study on Physics
%\ and Detectors for a Linear Electron-Positron Collider, held at Orsay,
%\ Lund, Frascati, Oxford, and Obernai from April 1998 until Oct 1999, to
%\ appear as DESY report 123F, ``Physics Studies for a Future Linear Collider''
%\ (R. Heuer, F. Richard, P. Zerwas, eds.)}
%\\
\\
$^{\ddag}$~{\small On leave from Departamento de F{\'\i}sica Te\'orica y del 
  Cosmos, Universidad de Granada, Fuentenueva s/n, E-18071 Granada, Spain.
\\
E-mails: {jillana@ifh.de}, {jack@ifh.de}, {riemann@ifh.de} }

\newpage
%  end of title page
% 
% \tableofcontents
% \listoftables
% \listoffigures
% 
% \newpage
% 
%=============================================== 
\section{\label{sec-intro}Introduction}
\setcounter{equation}{0}
%===============================================
With the Giga--$Z$ option of the Tesla project one may expect the
production of about $10^{9}$ $Z$ bosons at resonance \cite{Hawkings:1999ac}.
This huge rate, about a factor 100 higher than rates at LEP~1, allows 
one to study a number of problems with unprecedented precision. 
Among them is the search for lepton-flavour changes in $Z$ decays:
\ba
Z \to e\mu, \mu\tau, e\tau .
\label{eq1}
\ea
Non-zero rates are expected if neutrinos are massive and mix 
\cite{Pontecorvo:1957cp,Maki:1962mu,Pontecorvo:1968fh}.
Often one considers the branching ratio for the production of 
the following sum of charged states:
\ba
\label{lep-0}
{\rm BR}(Z\to l_1^{\mp}l_2^{\pm}) 
=
\frac{\Gamma(Z\to \bar l_1 l_2 + l_1 \bar l_2)}
%{\Gamma(Z\to{\rm all})}.
{\Gamma_Z}.
\ea
First predictions for flavour-changing $Z$ decays in the
framework of the Standard Model  
\cite{Glashow:1961ez,Weinberg:1967,Salam:1968rm}, using techniques
developed in \cite{'tHooft:1972fi}, were given in 
\cite{Riemann:1982rq,Mann:1984dvt,Ganapathi:1983xy,Clements:1983mk}.  
The best direct limits are obtained by searches at LEP~1 (95\% c.l.)
\cite{PDG:1998aa}: 
\ba
\label{lep-em}
{\rm BR}(Z\to e^{\mp}\mu^{\pm}) &<& 1.7 \times 10^{-6}~~ \cite{Akers:1995gz},
\\
\label{lep-et}
{\rm BR}(Z\to e^{\mp}\tau^{\pm}) &<& 9.8 \times 10^{-6} ~~
\cite{Adriani:1993sy,Akers:1995gz},
\\
\label{lep-tm}
{\rm BR}(Z\to \mu^{\mp}\tau^{\pm}) &<& 1.2 \times 10^{-5} ~~
\cite{Akers:1995gz,Abreu:1996mj}.
\ea
A careful analysis shows, taking into account realistic
conditions at future experiments,
that the sensitivities for the branching ratios 
could be improved considerably at the Giga--$Z$ \cite{Wilson:1998bb},
namely down to: 
\ba
\label{lep-emx}
{\rm BR}(Z\to e^{\mp}\mu^{\pm}) &<&  2 \times 10^{-9},
\\
\label{lep-etx}
{\rm BR}(Z\to e^{\mp}\tau^{\pm}) &<& f \times 6.5 \times 10^{-8},
\\
\label{lep-tmx}
{\rm BR}(Z\to \mu^{\mp}\tau^{\pm}) &<& f \times 2.2 \times 10^{-8},
\ea
with $f = 0.2 \div 1.0$.  

These numbers may be confronted with expectations derived from
the signals for $\nu_{\mu}-\nu_{\tau}$ oscillations in atmospheric
neutrino experiments  
 \cite{Fukuda:1994mc,Fukuda:1998mi,Fukuda:1998tw,Fukuda:1998ub% 
%indications,evidence
,Becker-Szendy:1995aa% indic.
,Allison:1997yb% confirm. of evidence
,Ambrosio:1998wu% confirm. of evidence
}.
They are at the 90\% c.l. compatible with the following parameter set
\cite{Mohapatra:1999em}:
\ba
\label{experimental1}
\Delta m_{\nu_{\mu}\nu_{\tau}}^2 &\simeq& (2 \div 8) \times 10^{-3} {\rm eV}^2,
\\
\label{experimental2}
\sin^2 (2\vartheta_{\mu\tau}) &\simeq& 0.8 \div 1.
\ea
There is also evidence for $\nu_{e}-\nu_{\mu}$ oscillations from solar
neutrino experiments  
\cite{Davis:1968cp,Cleveland:1998nv,Fukuda:1996sz,Hampel:1996qd,%
Abdurashitov:1996dp,Suzuki:1998}, being compatible with:
\ba
\label{mass-e-m}
\Delta m_{\nu_{e}\nu_{\mu}}^2 &\simeq& 10^{-10} \div 10^{-5} {\rm
eV}^2 .
\label{experimental3}
\ea
From reactor searches, there are no hints of $\nu_{e}-\nu_{\tau}$
oscillations \cite{Apollonio:1997xe,Apollonio:1999ae}.   
For more details see e.g. the review \cite{Mohapatra:1999em} and
references therein.

The good news from the evidences for neutrino oscillations is that
they suggest non-vanishing rates for reaction (\ref{eq1}).
The bad news is, that these rates are, if derived with
(\ref{experimental1})--(\ref{experimental3}) in the minimally extended Standard 
Model ($\nu$SM), 
extremely small\footnote{Our
estimate is in clear 
distinction to Eqn. (6) of \cite{Pham:1998xhp}, where from the data a
limit was derived which corresponds to 
${\rm BR}(Z\to\mu^{\mp}\tau^{\pm} )\approx {\cal O}(10^{-8} \div
10^{-5})$.
}:
% \lsim
\ba
\label{old-tm}
{\rm BR}(Z\to \mu^{\mp}\tau^{\pm}) &\sim&  10^{-54},
\\
{\rm BR}(Z\to e^{\mp} \mu^{\pm}) \sim
{\rm BR}(Z\to e^{\mp}\tau^{\pm}) &\lsim&  4 \times 10^{-60}.
\ea
This is derived in Section \ref{sec-SM} and is also in accordance
with older calculations
\cite{Riemann:1982rq,Mann:1984dvt,Riemann:1999ab}.
% extraordinarily small\footnote{Our
% estimate is in clear 
% distinction to Eqn. (6) of \cite{Pham:1998xhp}, where from the data a
% limit was derived which corresponds to 
% ${\rm BR}(Z\to\mu^{\mp}\tau^{\pm} )\approx {\cal O}(10^{-8} \div 10^{-5})$.}
% \cite{Riemann:1982rq,Mann:1984dvt,Riemann:1999ab}:
% \lsim
% \ba
% \label{old-tm}
% {\rm BR}(Z\to \mu^{\mp}\tau^{\pm}) &\sim&  10^{-54},
% \\
% {\rm BR}(Z\to e^{\mp} \mu^{\pm}) \sim
% {\rm BR}(Z\to e^{\mp}\tau^{\pm}) &\lsim&  4 \times 10^{-60}.
% \ea

How do these small numbers arise?
In Born approximation,
the lepton-flavour changing $Z$ decay into two charged leptons is forbidden
in the $\nu$SM due to the GIM mechanism \cite{Glashow:1970st}.
However, it may take place if $n$ types of neutrinos have masses $m_i$
and mix with each other (with mixing matrix ${\bf V}_{ij}$),  
i.e. if symmetry eigenstates and mass eigenstates are
different in the lepton sector.
Then, the virtual exchange of these neutrinos produces an effective
lepton-flavour changing vertex and the corresponding  
branching ratio has the following structure:
\ba
\label{lep-1}
{\rm BR}(Z\to l_1^{\mp}l_2^{\pm})
&=& \frac{\alpha^3}{192\pi^2 s_W^6c^2_W}\frac{M_Z}{\Gamma_Z}
~|{\cal V}(M^2_Z)|^2
\approx 10^{-6}~|{\cal V}(M^2_Z)|^2.
\ea
%Further, as will be shown in Section \ref{sec-SM}, it is:
The form factor ${\cal V}(Q^2)$ depends on the details of the interaction:
\ba
\label{order-V}
 {\cal V}(Q^2) &=& \sum_{i=1}^n {\bf V}_{l_1i} {\bf V}_{l_2i}^* ~
V(m_i^2/M_W^2). 
\ea
The vertex function $V$ was calculated in 1982/83 
independently by three groups  
\cite{Riemann:1982rq,Mann:1984dvt,Ganapathi:1983xy,Clements:1983mk}
for sequential Dirac particles,
and in a more general context later \cite{Korner:1993an,Ilakovac:1995kj}.

The function $V$ depends quadratically on the mass
{\em both} in the small neutrino mass limit 
\cite{Mann:1982xw,Riemann:1982rq}
{\it and} in the large neutrino mass limit 
\cite{Riemann:1982sx,Mann:1984dvt,Ganapathi:1983xy,Clements:1983mk}:
\ba
\label{smallD}
V(\lambda_i \ll 1) - V(0) &\approx&
%(1.25 + 1.03 \times i) + (2.5623 - 2.2950 \times i) \frac{m_i^2}{M_W^2}
%(1.25 + 1.03 \times i) + (2.56 - 2.30 \times i) ~\lambda_i
(2.56 - 2.30 \times i) ~\lambda_i
+~ {\cal O}\left(\lambda_i^2\ln\lambda_i\right) ,
\\
\label{bigD}
V(\lambda_i \gg 1) - V(0) &\approx&  \frac{1}{2} ~\lambda_i 
+ {\cal O}(\ln\lambda_i) ,
\ea
with
\ba
\label{delta}
\lambda_i &=& \frac{m_i^2}{M_W^2}.
\ea
The constant terms are not shown in (\ref{smallD}), (\ref{bigD}) since they
drop out in (\ref{order-V}) due to the unitarity of the mixing matrix,
and the branching ratio becomes
proportional to the fourth power of the neutrino massses.
It is this behaviour which makes the expected rates so extremely
small for the experimentally evidenced tiny neutrino masses.
For values of $m_i$ of the order of $M_W$, the vertex  $V$ is of the order
one, and could become large if the $m_i$ would be much bigger 
than $M_W$.

\bigskip

The evidence of tiny neutrino masses may also be
indicative for a mechanism which produces at the same time 
very large masses. 
Heavy neutrinos are expected by some GUTs \cite{Langacker:1981js} and
string-inspired models 
\cite{Witten:1986bz,Mohapatra:1986bd,Hewett:1989xc}, and are suggested by
the seesaw mechanism
\cite{Yanagida:1980xy,Gell-Mann:1980vs,Mohapatra:1980ia}.
Therefore, the above observations motivate us to have a closer look at
the prospects of observing lepton-flavour changes with the Tesla Linear
Collider.
%{\em
For the Giga--$Z$ option, 
it might not be sufficient to apply the well-known and simple 
approximations for large masses $m_i$, but also the medium- or 
even small-mass cases may be of experimental interest.
%}    

\bigskip

To be concrete, we will explore the following scenarios: 

\begin{itemize}

\item[(i)] The $\nu$SM. We treat the known light neutrinos ($\nu_e,\nu_\mu,
\nu_\tau$) as {\em massive Dirac} particles.
% (there is no difference
%between Dirac and 
%Majorana character for small masses 
%\\
%{\tt(this is to be checked later)}. 
%\\
Individual lepton 
numbers $L_e, L_\mu, L_\tau$ are not conserved any more. 
The lepton sector is then in exact analogy to the quark sector. 
As a by-product, the $Z$ decay amplitude into two quarks of 
different flavours can be read off from our general expressions.
% and
%some useful checks and 
%comparisons can be performed.

\item[(ii)] 
The $\nu$SM {\em sequentially extended} with {\em one} heavy 
{\em ordinary Dirac} neutrino.  
This case implies the existence of a heavy charged lepton as well%
%, in order to
%keep the  
%SU(2)$\otimes$U(1) gauge symmetry,
\footnote{A fourth generation of quarks is
also needed to keep the theory anomaly free.}.
% which does not get involved 
%anyway in the process $Z\to \bar{l}_1 l_2$ to one loop. 
It is not a very
favoured scenario but we consider it as a simple application of the
expressions of case (i) for heavier neutrinos. Again, total lepton
number $L$ is conserved.

\item[(iii)]
The $\nu$SM {\em extended} with {\em two} heavy {\em right-handed singlet 
Majorana} neutrinos. 
Not only individual, but also total $L$ is, in general, not conserved
since the presence of Majorana mass terms involves mixing of neutrinos and 
their charge-conjugate partners (antineutrinos), with opposite fermion-number. 
For two equal and heavy masses this case reduces to the addition of one heavy 
{\em singlet Dirac} neutrino \cite{Ilakovac:1995kj}. In this latter case $L$ is
recovered \cite{Wyler:1983dd}. 

\end{itemize}

% A heavy 
%sector is necessary to achieve sizeable lepton-flavour changing effects in
%$Z$  
%decays. 
%Nevertheless, we are in this work 

With our numerical estimates we will be
as model-independent as possible and will assume 
no constraints on neutrino masses or mixings, except for the ones imposed by the 
unitarity of the leptonic and neutrino mixing matrices, by the present bounds 
on lepton universality, CKM unitarity, and the measured $Z$ boson invisible
width  
\cite{Langacker:1988ur,Nardi:1992rg,Nardi:1994iv,%
Bergmann:1998rg,Bhattacharya:1995bj,Kalyniak:1997cs}, and by 
oscillation experiments (see \cite{Mohapatra:1999em} for a review).

In the following sections, we will discuss the predictions of the
three scenarios
for the lepton-number changing $Z$ decay.
In Appendices, the generalization of Lagrangian and Feynman rules of the
Standard Model to the general case of Dirac and Majorana masses is explained,
experimental limits on neutrino mixings and masses are quoted, 
and the calculation of the vertex function is sketched.
%=======================================================================
\section{\label{sec-SM}
%Predictions in 
Predictions from the $\nu$SM}
\setcounter{equation}{0}
%=======================================================================
\begin{figure}
D1:\put(10,-30){\epsfig{file=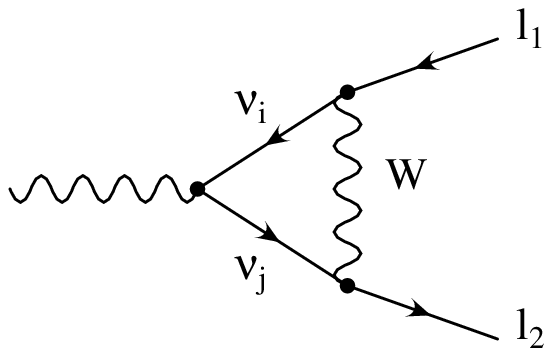,angle=0,width=0.25\linewidth}}
\hspace{4.5cm}
D2:\put(10,-30){\epsfig{file=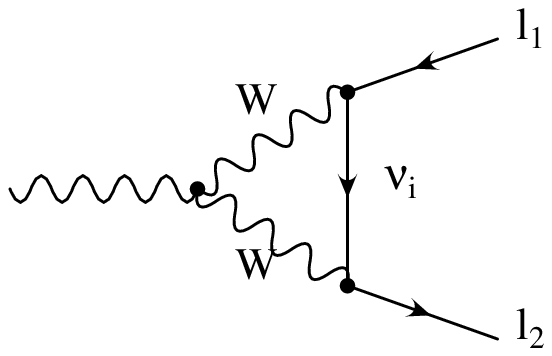,angle=0,width=0.25\linewidth}}
\hspace{4.5cm}
D3:\put(10,-30){\epsfig{file=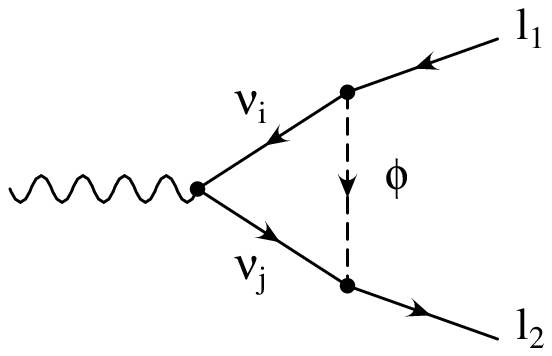,angle=0,width=0.25\linewidth}} \\
D4:\put(10,-30){\epsfig{file=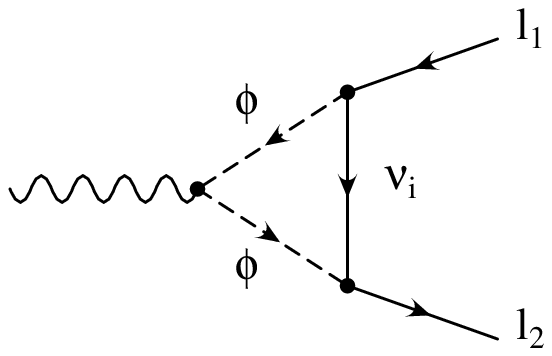,angle=0,width=0.25\linewidth}}
\hspace{4.5cm}
D5:\put(10,-30){\epsfig{file=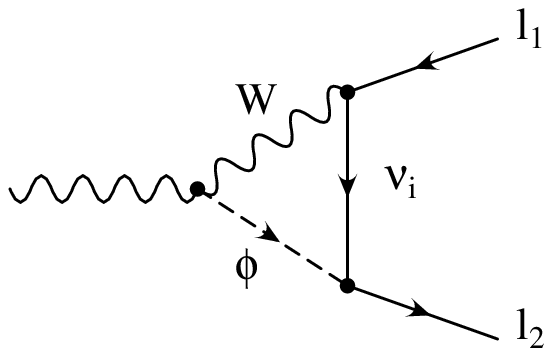,angle=0,width=0.25\linewidth}}
\hspace{4.3cm} + crossed \\
D$\Sigma$:
 \put(10,-23){\epsfig{file=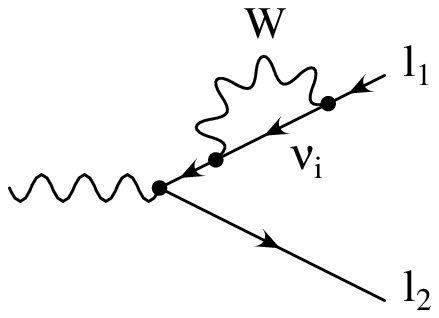,angle=0,width=0.2\linewidth}}
\hspace{3.3cm}
+\put(10,-23){\epsfig{file=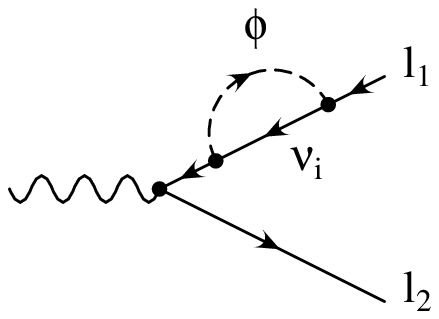,angle=0,width=0.2\linewidth}}
\hspace{3.3cm}
+\put(10,-36){\epsfig{file=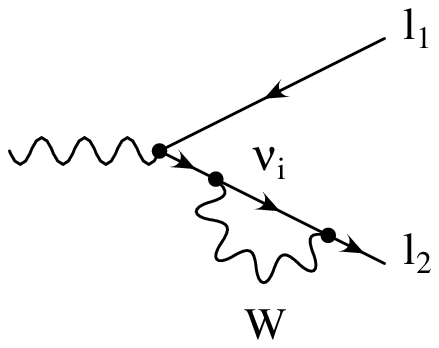,angle=0,width=0.2\linewidth}}
\hspace{3.3cm}
+\put(10,-38){\epsfig{file=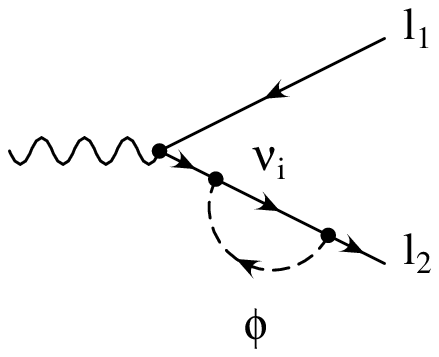,angle=0,width=0.2\linewidth}}
\caption{
\label{fig-vertex}
\it 
Feynman diagrams for the lepton-flavour changing $Z$ decay. 
In the case of virtual, ordinary Dirac neutrinos, 
the $Z\nu_i\nu_j$ vertices in D1 and D3 
are diagonal and the analogous quark-flavour-changing process can be
obtained by replacing $l_k$ by down-quarks and $\nu_i$ by up-quarks. 
%For Majorana neutrinos, the arrows are just arbitrary orientations 
%rather than fermion-number flows (see discussion in Appendix ?).
\label{fig1}
}
\end{figure}

The amplitude for the decay of a $Z$ boson into two charged leptons with 
different flavour, $l_1$ and $l_2$, is given in a self-explanatory notation 
by:
\beq
{\cal M}=-
\frac{ig\alpha_W}{16\pi c_W}
%\frac{ig^3}{64\pi^2 c_W}
\ {\cal V}(Q^2)\ 
\varepsilon^\mu_Z \bar u_{l_2}(p_2)\gamma_\mu(1-\gamma_5) u_{l_1}(-p_1),
\label{ampli}
\eeq
where 
\ba
\label{l22}
\alpha_W \equiv  \frac{\alpha}{s_W^2}, %,g^4/4\pi
\ea
and the form factor ${\cal V}$ depends on 
$Q^2=(p_2-p_1)^2$ and can be written as: 
\ba
\label{l23}
{\cal V}(Q^2) &=& \sum^3_{i=1} {\bf V}_{l_1 i} {\bf V}^{^*}_{l_2 i}
~V(\lambda_i),
\\
\label{l24}
V(\lambda_i) &=&
\left[ 
v_W(i)+v_{WW}(i)+v_{\phi}(i)+v_{\phi\phi}(i)+v_{W\phi}(i)+v_{\Sigma}(i)
\right],
\ea
with ${\bf V}_{i j}$ being the leptonic CKM mixing matrix. 
In general, there are besides the vector and axial-vector couplings
$f_V$ and $f_A$ in 
(\ref{ampli}) also contributions of the $f_S, f_P, f_M, f_E$ types, but for 
the production of on-shell fermions (with their masses being neglected) 
they vanish here.
Further, it is $f_V = f_A = {\cal V}(Q^2)$ due to the presence of 
$W$ bosons coupling only to left-handed fermions.
%virtual left-handed $W^{\pm}$ bosons.
The contributions
from the one-loop diagrams of Figure~\ref{fig1} depend on
$\lambda_i$ and additionally on  
\ba
\label{deltaq}
\lambda_Q &=& \frac{Q^2}{M_W^2},
\ea
which on the $Z$ boson mass shell becomes
\ba
\label{deltaz}
\lambda_Z &=& \frac{M_Z^2}{M_W^2}=\frac{1}{c^2_W} \approx 1.286.
\ea
In terms of the usual vector and 
axial-vector couplings,
\ba
v_i &=& I^{i_L}_3 - 2 Q_i s_W^2 = I^{i_L}_3(1-4 s^2_W |Q_i|),
\\
a_i &=& I^{i_L}_3,
\ea
the individual contributions are:
\begin{itemize}
\item
From vertex diagrams: 
\beq
\label{D1}
{\rm D1:} \ \ \ v_W(i) & = & 
-(v_i+a_i)\ \left[ \lambda_Q\ (C_{0}+C_{11}+C_{12}+C_{23})-2C_{24}+1\right] 
\nonumber\\
                  &   & -(v_i-a_i)\ \lambda_i\ C_{0},
\\ 
{\rm D2:}\  v_{WW}(i)  & = & 2c^2_W\ (2I^{i_L}_3) \left[ \lambda_Q\ 
({\bar C}_{11}+{\bar C}_{12}+{\bar C}_{23})-6{\bar C}_{24}+1\right],
\\  
\label{D3}
{\rm D3:} \ \ \ \  v_{\phi}(i)  & = &-(v_i+a_i)\ 
\displaystyle\frac{\lambda^2_i}{2}\ C_0 
\nonumber\\
                  &   & -(v_i-a_i)\ \displaystyle\frac{\lambda_i}{2}\ 
          \left[\lambda_Q\ C_{23} - 2 C_{24} + \displaystyle\frac{1}{2} \right],
\\  
{\rm D4:}\ \ \ v_{\phi\phi}(i)  & = &-(1-2s^2_W)\ (2I^{i_L}_3)\ \lambda_i\
{\bar C}_{24},
\\  
{\rm D5:}\ \  v_{W\phi}(i)  & = & -2s^2_W\ (2I^{i_L}_3)\ \lambda_i\ {\bar C}_0;
\eeq

\item
From self-energy corrections to the external fermion lines:
\beq
{\rm D}\Sigma:\ v_{\Sigma}(i)  & = & 
\displaystyle\frac{1}{2}(v_i+a_i-4c^2_W a_i)\left[(2+\lambda_i)B_1 +1 \right].
\eeq

\end{itemize}
The one-loop tensor integrals 
$C_0$, ${\bar C}_0$, $C_{ij}$, ${\bar C}_{ij}$, and $B_1$ 
are defined in Appendix \ref{app-vertex}.
With our numerical results for the Dirac case, we rely on two
calculations, an old one 
\cite{Riemann:1982rq,Mann:1984dvt} and also this new, completely 
independent one.
In the latter,
the numerical evaluation of the tensor integrals is performed with the
help of the computer program package {\tt LoopTools}
\cite{Hahn:1999wr,Hahn:1999mt}. 

\begin{figure}[th]
\begin{center}
\epsfig{file=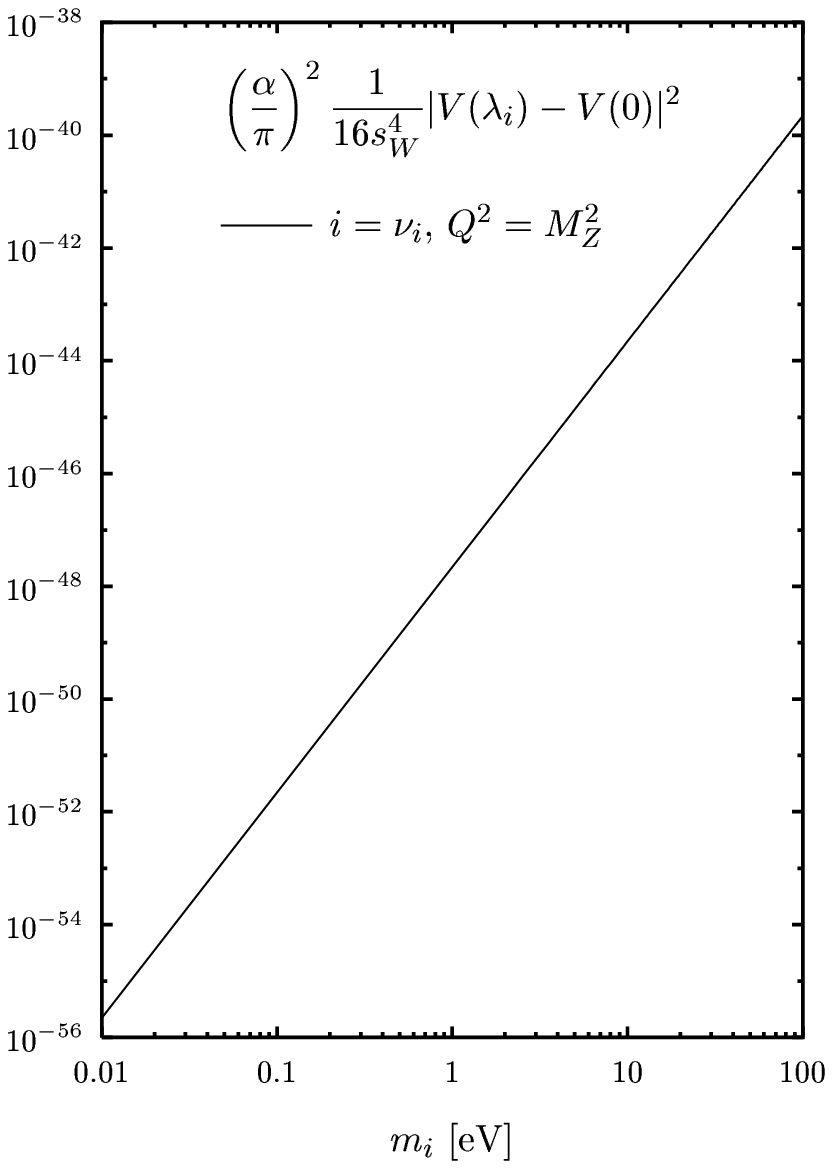,angle=0,width=0.49\linewidth}
\epsfig{file=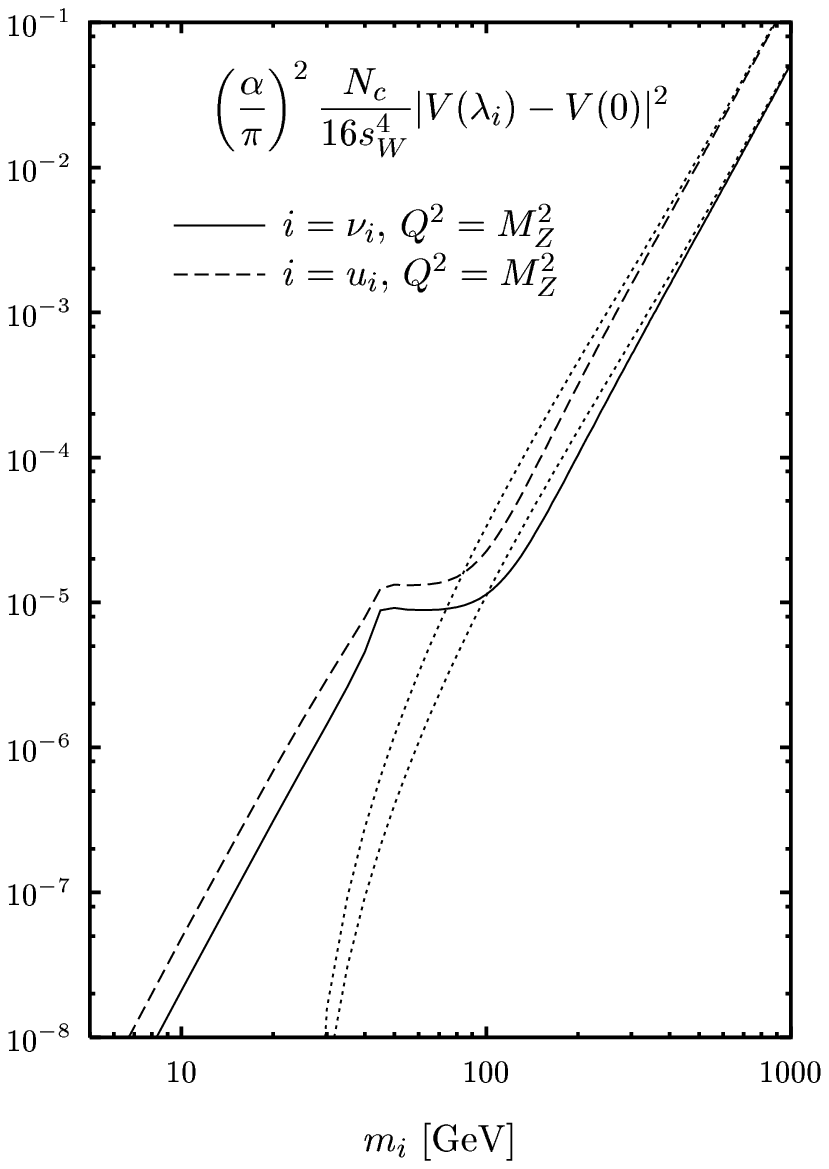,angle=0,width=0.49\linewidth}
\end{center}
\caption{
\label{figure-dirac}
{\it 
Contribution of one neutrino generation to
the ratio $B_Z$ of Eqn.~(\ref{bsubz}) in the small and large mass regions
for virtual, ordinary Dirac neutrinos and the analogous quark case. 
We set
${\bf V}_{l_1 i}{\bf V}^{*}_{l_2 i}=1$ here; 
dotted lines correspond to $Q^2=0$.}
}
\end{figure}

Numerical results are shown in Figure \ref{figure-dirac}.
The quantity presented is related to a branching ratio definition
often used in the literature:
\ba
\label{bsubz}
B_Z &\equiv& \frac{\Gamma(Z\to f_1^{\mp}f_2^{\pm}
%\bar l_1 l_2 + l_1 \bar l_2
)}
                          {2 \times \Gamma(Z\to \bar \nu_l \nu_l)}
= \left(\frac{\alpha}{\pi}\right)^2\frac{N_c}{16s^4_W}
~|{\cal V}(M^2_Z)|^2, 
\ea
with $N_c$ as colour factor.
Using 
\ba
\label{gammann}
\Gamma(Z\to {\bar \nu}_l \nu_l) &=&
\frac{\alpha_W}{24c_W^2} M_Z,
\ea
%\Gamma(Z\to {\bar f} f) &=&
% \frac{\alpha_W}{24c_W^2} M_Z (1-4s_W^2|Q_f|+8s_W^4|Q|^2),
we get a useful relation to the branching ratio introduced in
(\ref{lep-0}) and (\ref{lep-1}): 
\ba
\label{relat}
{\rm BR}(Z\to f_1^{\mp}f_2^{\pm}
%\bar l_1 l_2 + l_1 \bar l_2
)
&=&
\frac{2 \times \Gamma(Z\to {\bar \nu}_l \nu_l)}{\Gamma_Z} ~B_Z 
%.\nl
= 0.1333 ~ B_Z .
\ea

%\subsection{Decay width and various branching ratios}
%=========================================================================
%=========================================================================

Figure \ref{figure-dirac} shows the contribution from one neutrino generation to the
branching ratio as a function of the neutrino mass {\em without} the
influence of the mixing matrix elements.
We choose two interesting mass regions, namely that corresponding to
the findings of neutrino oscillation searches (Figure
\ref{figure-dirac}(a)) and also that with 
potential predictions in the reach of the Giga--$Z$ 
(Figure \ref{figure-dirac}(b)).
The latter figure nicely agrees with the earlier calculations
\cite{Mann:1984dvt,Ganapathi:1983xy,Clements:1983mk}, 
and the small mass limit with \cite{Riemann:1982rq}.  
The dotted lines in Figure \ref{figure-dirac}(b) correspond to the
approximation $\lambda_Z=0$: 
the large mass limit is reproduced quite well, while the small mass
limit differs from the correct result. 
This is discussed in Appendix \ref{mapra}.

We will now estimate the branching ratio under the assumption that
there are three generations of light neutrino flavours with a unitary
mixing matrix ${\bf V}$ as evidenced by experiment.
%The experimental input to be used is given in Appendix \ref{app-exp}.
% from J. Ellis, hep-ph/9907458
The general form of this matrix may be chosen to be \cite{Maki:1962mu}:
\ba
%\begin{gather*}
{\bf V}_{l\nu} 
&=& 
\begin{pmatrix}
{\bf V}_{e1}      &  {\bf V}_{e2}       &  {\bf V}_{e3} 
\\ 
{\bf V}_{\mu 1}   &  {\bf V}_{\mu 2}    & {\bf V}_{\mu 3}
\\ 
{\bf V}_{\tau 1}  &  {\bf V}_{\tau 2}   &  {\bf V}_{\tau 3}  
\end{pmatrix}
\nonumber\\
&=&
% remark of JI:
% standard PDG: the phase delta together with s23 instead of with s12
\begin{pmatrix}
c_{12}c_{13} & c_{13}s_{12}  & s_{13}  
\\
 -c_{23}s_{12}e^{i\delta}-c_{12}s_{13}s_{23} 
& c_{12}c_{23}e^{i\delta}-s_{12}s_{13}s_{23}  
& c_{13}s_{23}  
\\ 
  s_{23}s_{12}e^{i\delta}-c_{12}c_{23}s_{13} 
&-c_{12}s_{23}e^{i\delta}-c_{23}s_{12}s_{13} 
& c_{13}c_{23}  
\end{pmatrix}
\begin{pmatrix}
e^{i\alpha} & 0 & 0
\\
0 & e^{i\beta} & 0 
\\
0 & 0 & 1
\end{pmatrix}
%\end{gather*}
\nl
\ea

Here, we have three mixing angles and one CP-violating phase as in the
quark CKM case, plus two CP-violating phases $\alpha,\beta$ if neutrinos are
Majorana particles (they are {\em strictly} neutral 
so that less phase factors may be `eaten' by redefining complex
fermion fields).   
Current data suggest the following form of this matrix:
\ba
\label{mixmat}
{\bf V}_{l\nu} 
=
\begin{pmatrix}
c_{12} & s_{12} & 0
\\
-\frac{1}{\sqrt{2}}s_{12} & \frac{1}{\sqrt{2}}c_{12} & \frac{1}{\sqrt{2}}
\\
\frac{1}{\sqrt{2}}s_{12} & -\frac{1}{\sqrt{2}}c_{12}&\frac{1}{\sqrt{2}}
\end{pmatrix},
\ea
where we have assumed \cite{Peccei:1998es} 
%quoted from Pham
$\alpha = \beta = \delta =0$, 
extracted from \cite{Apollonio:1997xe,Apollonio:1999ae} the 
$s_{13}=0$, and further assumed $s_{23} = 
1/\sqrt{2}$ (corresponding to maximal mixing) and left the
$\theta_{12}$ free.

With %these assumptions and with
\ba
\label{mw}
M_W &=& 80.41~ {\rm GeV},
\\
\label{mz}
M_Z &=& 91.187~  {\rm GeV},
\\
\label{gamz}
\Gamma_{Z} = \Gamma(Z\to{\rm all}) &=& 2.49~  {\rm GeV},
\ea
we get after trivial calculations:
\ba
{\rm BR}(Z\to l_1^{\mp}l_2^{\pm}
%\bar l_1 l_2 + l_1 \bar l_2
)
 &=&  
\frac{\alpha_W^3M_Z}{192\pi^2c_W^2\Gamma_Z}
\Bigl| {\bf V}_{l_11}{\bf V}_{l_21}^*[V(\lambda_1) - V(0)]
\nl 
&&+~{\bf V}_{l_12}{\bf V}_{l_22}^*[V(\lambda_2) - V(0)]
+{\bf V}_{l_13}{\bf V}_{l_23}^*[V(\lambda_3) - V(0)]
\Bigr|^2,
\nl
%|a_1|^2
\ea
with 
\ba
\frac{\alpha_W^3M_Z}{192\pi^2c_W^2\Gamma_Z} = 1.127 \times 10^{-6}.
\ea
For small neutrino masses, we show in Section \ref{sec-marie}:
\ba
V(\lambda_i) - V(0) = a_1(\lambda_{Z}) ~\lambda_i.
\ea
The resulting branchings are, without approximations yet, but using
the information from three-generation unitarity:
\ba
{\rm BR}(Z\to l_1^{\mp}l_2^{\pm}
%\bar l_1 l_2 + l_1 \bar l_2
)
 &=&  \frac{\alpha_W^3M_Z}{192\pi^2c_W^2\Gamma_Z}
~ |a_1|^2 ~ 
\left|
{\bf V}_{l_11}{\bf V}_{l_21}^*\lambda_{12}
-
{\bf V}_{l_13}{\bf V}_{l_23}^*\lambda_{23} 
\right|^2,~~
\ea
with
\ba
\label{deflambda}
\lambda_{ij} &=& |\lambda_i - \lambda_j|
\ea
and the $a_1$ is given in (\ref{a1}):
\ba
|a_1|^2 = 11.832.
\ea
There are two different cases to be considered when using now the specific 
mixing matrix (\ref{mixmat}):
\ba
{\rm BR}(Z\to e^{\mp} \mu^{\pm}
%\bar e \mu + e \bar \mu
)
\approx
{\rm BR}(Z\to e^{\mp} \tau^{\pm})
&\approx&
1.333 \times  10^{-5} ~\frac{c_{12}^2s_{12}^2}{2} ~\lambda_{12}^2,
\ea
and
\ba
{\rm BR}(Z\to \mu^{\mp} \tau^{\pm})
&=&
1.333 \times  10^{-5} ~\frac{1}{4}~ 
\left|s_{12}^2  \lambda_{12} - \lambda_{23} \right|^2.
\ea
From the mass estimate (\ref{experimental1}) we get
as additional input from atmospheric neutrino studies:
\ba
\lambda_{23} \approx
\frac{(2\div8) \times 10^{-3} {\rm eV}^2} {M_W^2} = (3\div12) \times 10^{-25},
\ea
and from solar neutrino searches, (\ref{mass-e-m}):
\ba
\lambda_{12} 
\approx 
%\frac{1}{2} \times \left(10^{-7} \div 10^{-2} \right) \lambda_{23}= 
1.5 \times \left(10^{-32} \div 10^{-27} \right).
%<< \lambda_{23} . 
\ea
It is easy now to see that the expected rates for the lepton-flavour
changing $Z$ decays are limited to:
\ba
\label{maxzdec}
{\rm BR}(Z\to e^{\mp} \mu^{\pm}
%\bar e \mu + e \bar \mue^{\pm} \mu^{\mp}
)
\approx
{\rm BR}(Z\to e^{\mp}\tau^{\pm}
%\bar e \tau + e \bar \tau
)
&<&
%1.333 \times  10^{-5} \times \frac{1}{8} \times \left(
%\frac{1}{2}10^{-2}\right)^2  
%\times \left( 3.018 \times  10^{-25}\right)^2
3.75  \times  10^{-60}, 
%\nl
%&=& ...
\\
{\rm BR}(Z\to \mu^{\mp}\tau^{\pm}
%\bar \mu \tau + \mu \bar \tau
)
&\approx&
(3\div48) \times  10^{-55}.
\ea
In (\ref{maxzdec}), we assumed arbitrarily a maximal mixing
$s_{12}=1/\sqrt{2}$.
These rates are extremely small. In fact, we will neglect the effects
of the light neutrino sector in the next sections, where we extend the
$\nu$SM to accommodate heavy neutrinos, taking massless the known ones. 
%tord 091299
%=======================================================================
\section{\label{sec-sequent}
%Scenario (ii)
Predictions from the $\nu$SM plus One Heavy Dirac Neutrino
}
\setcounter{equation}{0}
%======================================================================
Here, the only effective difference to the case  before is the
existence of a fourth generation with a sequential Dirac neutrino of mass
$m_N$. 
In this case, the branching ratio gets:
\ba
{\rm BR}(Z\to l_1^{\mp}l_2^{\pm}
%\bar l_1 l_2 + l_1 \bar l_2
)
 &=&  
\frac{\alpha_W^3M_Z}{192\pi^2c_W^2\Gamma_Z}
~\left| {\bf V}_{l_1N}{\bf V}_{l_2N}^*\right|^2
~\left|V(\lambda_N) - V(0)\right|^2,
\ea
with $V(\lambda_N)$ given in (\ref{l24}).
The numerical results depend crucially on the mixing between 
the light and heavy leptons.
An optimistic assumption would be maximal mixing:
\ba
\label{optim}
\left| {\bf V}_{l_1N}{\bf V}_{l_2N}^*\right|^2
= \left(\frac{1}{\sqrt{2}}\right)^4 = \frac{1}{4}.
\ea
This is of course unrealistic.
Stringent (though indirect) limits may be derived
from the analysis of flavour-diagonal 
reactions as advocated e.g. in
\cite{Nardi:1992rg,Nardi:1994iv,Bergmann:1998rg} as well as from the
lepton flavour-changing process $\mu\to e\gamma$ 
\cite{Langacker:1988up,Tommasini:1995ii}.  
A short summary may be found in Appendix \ref{app-exp}.
There the matrix ${\bf B}$ is, for this particular case of heavy
Dirac neutrinos, ${\bf B}={\bf V}$. 

In order to be definite, we show in Figure \ref{figure-bergmt}a,
solid line, the predictions for  ${\rm BR}(Z\to \mu^{\pm} \tau^{\mp})$
and  assume the upper limit of the mixings allowed from 
(\ref{b11})--(\ref{b13}) and (\ref{fd2}): 
\ba
\left| {\bf V}_{\mu N}{\bf V}_{\tau N}^*\right|^2 < 1.5 \times 10^{-4}.
\label{mutau}
\ea
For the other two lepton flavour-changing $Z$ decay channels, the corresponding
graphs scale simply in accordance with the ratios of the mixing matrix elements:
\ba
\left| {\bf V}_{e N}{\bf V}_{\tau N}^*\right|^2 &<& 1.9 \times 10^{-4},
\label{etau}
\\\left| {\bf V}_{e N}{\bf V}_{\mu N}^*\right|^2 &<& 1.2 \times
10^{-4}.
\label{emu}
\ea
These coupling factors are to be  
compared to (\ref{optim}):
we observe a suppression of the expected branching ratio (for a given mass
$m_N$) by more than three orders of magnitude. 

As mentioned in Appendix \ref{app-exp}, these bounds on light-heavy mixings 
from flavour-diagonal processes are improved by flavour-changing processes 
involving the first two lepton families. In fact, from \cite{Brooks:1999pu}
%\cite{PDG:1998aa}:
\ba
%{\rm BR}(\mu\to e\gamma)<4.9\times10^{-11},
{\rm BR}(\mu\to e\gamma)<1.2\times10^{-11},
\ea
one obtains, for heavy neutrinos, the following (nearly) mass-independent limit
\cite{Langacker:1988up,Tommasini:1995ii}:
\ba
%\left| {\bf V}_{e N}{\bf V}_{\mu N}^*\right|^2 &<&5.76\times10^{-8},
\left| {\bf V}_{e N}{\bf V}_{\mu N}^*\right|^2 &<&1.4\times10^{-8},
\ea
much more stringent than (\ref{emu}).

The Giga--$Z$ discovery range is indicated in Figure \ref{figure-bergmt}a,
using the maximum values of the mixings allowed by (\ref{mutau}). 
The range will be limited to the large neutrino mass limit
if we believe in the relevance of the above mixing bounds. 
Then, the approximations of 
%section \ref{app-bigm}, combined with (\ref{a0ana}), (\ref{a0}) 
(\ref{d40})
apply.
Numerically, this means for $\lambda_Z = 1.286$:
\ba
\label{eq3-6}
\left|V(\lambda_N) - V(0)\right|^2
&=&
\frac{1}{4} ~\lambda_N^2 + 1.44~ \lambda_N\ln\lambda_N
-3.49 ~\lambda_N 
+ 2.07 ~\ln^2\lambda_N
+ {\cal O}(\ln\lambda_N).
\nl
\ea
If neutrinos with a mass of several hundred GeV or more would exist,
there is a good chance to observe some effect of them from the $Z$ decays 
under study. Indeed one cannot constrain heavy neutrino masses
from these processes since they also depend on the mixings, and vice versa. 
The fact that
in Figure \ref{figure-bergmt}a the discovery reach of LEP or Giga--$Z$
cuts the curves means: if no event is observed, mixings must be smaller 
than the ones employed (we took present upper bounds) for neutrino masses 
above the intersection point; or if some effect was observed then 
the plot would provide a lower bound for the heavy neutrino mass. 
In fact, it is roughly only the product 
$m^4_N\left| {\bf V}_{l_1 N}{\bf V}_{l_2 N}^*\right|^2$ which can be 
constrained, assuming only one heavy Dirac neutrino.

%=======================================================================
\section{\label{sec-Majorana}
%Scenario (iii)
Predictions from the $\nu$SM plus Two Heavy 
\\
Right-Handed Singlet Majorana 
Neutrinos}
\setcounter{equation}{0}
%=======================================================================
Some basic features of Lagrangians with Majorana mass terms 
and their relations to the simpler Dirac case are summarized in Appendix
\ref{app-feynman}.
For a derivation of Feynman rules with Majorana particles 
\cite{Majorana:1937vz} 
we refer to \cite{Schechter:1980gr,Gluza:1992wj,Denner:1992vz}. 
The couplings of the virtual neutrinos to the
$W$ bosons are left-handed, $v_i+a_i=1$, $v_i-a_i=0$, and
$2I^{i_L}_3=1$,
while those to the $Z$ boson and to the Higgs particles are
non-diagonal and contain right-handed admixtures.
This may be seen from (\ref{lw})--(\ref{lh}).
The amplitude for the decay (\ref{eq1}) 
is again given by Eqn.~(\ref{ampli}), but  
this time the form factor ${\cal V}_M$ is non-diagonal not only in the
external charged leptons, but also in the virtual neutrinos due to the  
$Z\nu_i\nu_j$ coupling, see (\ref{lz}):
\ba
\label{calvm}
{\cal V}_M(Q^2)&=&\sum^{n_G+n_R}_{i,j=1} {\bf B}_{l_1 i} {\bf B}^{*}_{l_2 j} 
~V(i,j),
\\
V(i,j) &\equiv& V(\lambda_i,\lambda_j,{\bf C}_{ij}) 
\nl
&=&
\left[ 
v_W(i,j)+\delta_{ij}v_{WW}(i)+v_{\phi}(i,j)+\delta_{ij}v_{\phi\phi}(i)
+\delta_{ij}v_{W\phi}(i)+\delta_{ij}v_{\Sigma}(i)
\right].
\nl
\ea
%=========================================================================
\begin{figure}[t]
\begin{center}
\epsfig{file=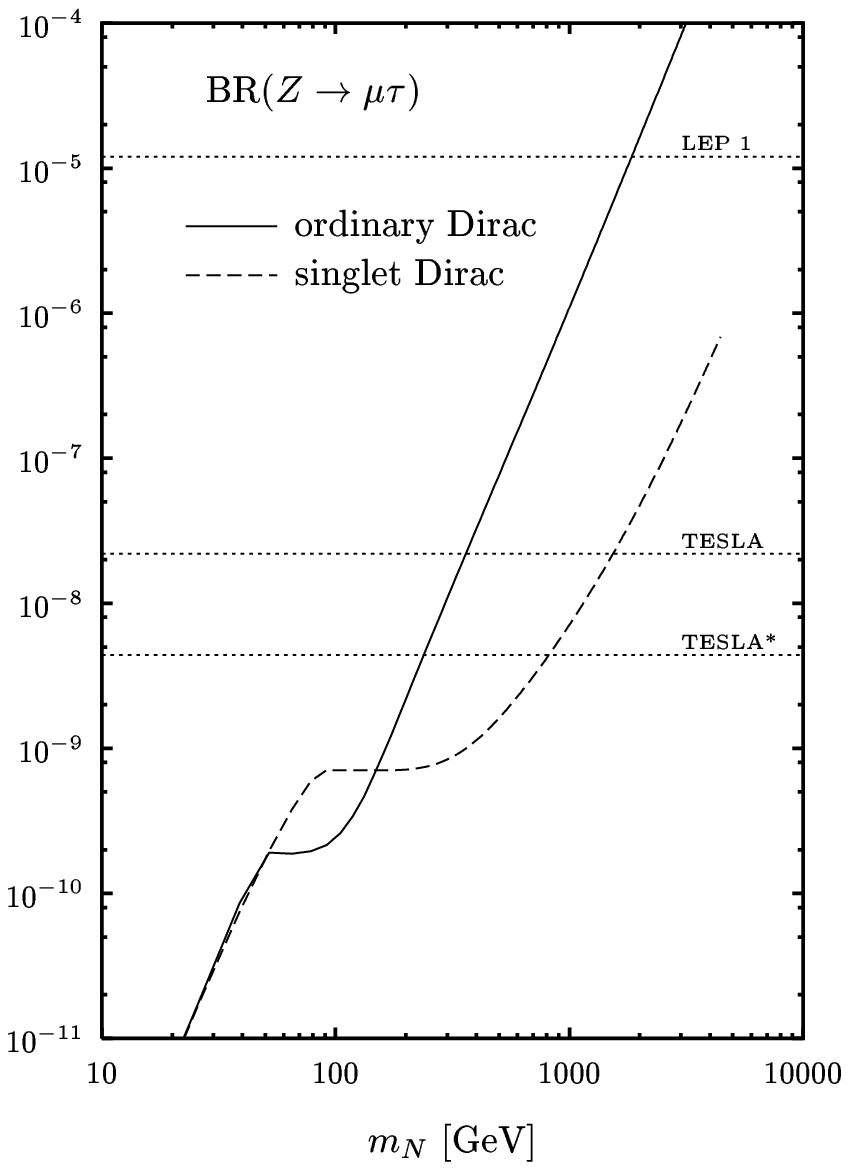,angle=0,width=0.49\linewidth}
\epsfig{file=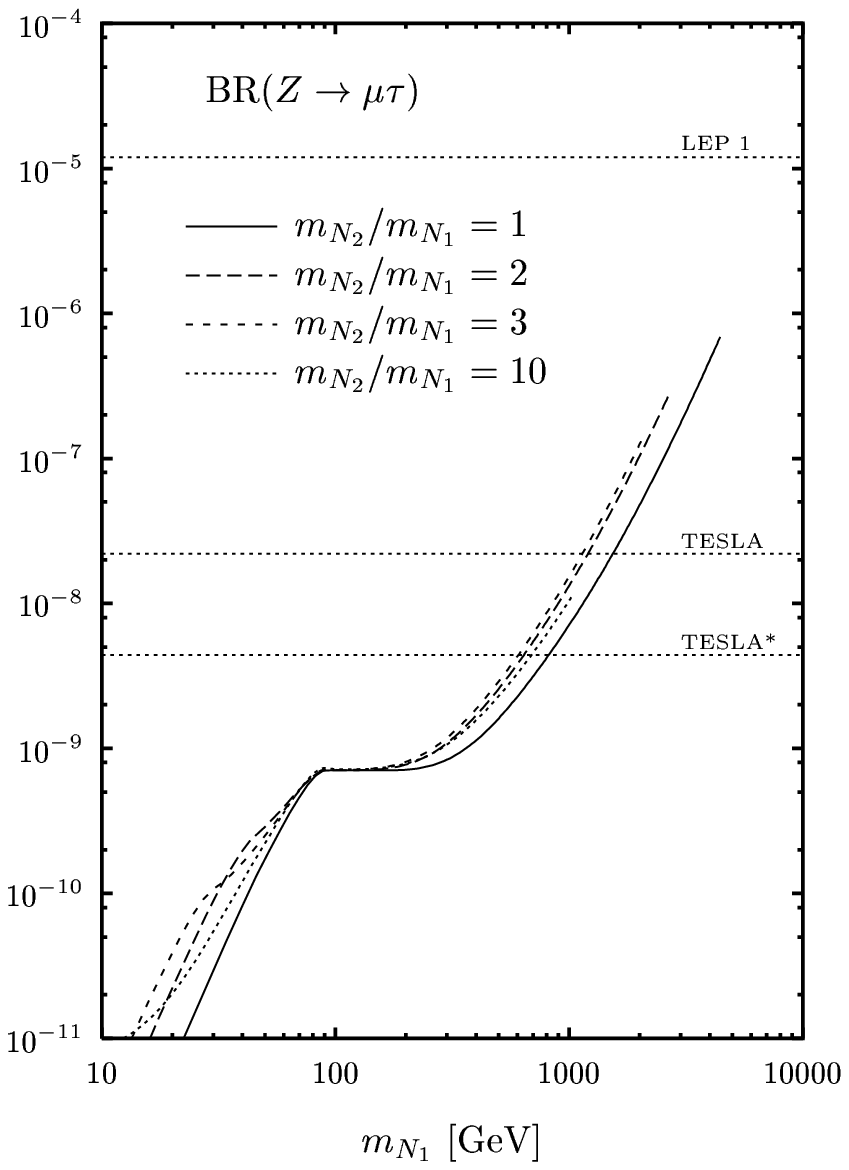,angle=0,width=0.49\linewidth}
\end{center}
\caption{
\label{figure-bergmt}
{\it 
Predictions for the branching fraction ${\rm BR}(Z\to \bar \mu \tau +
\mu \bar \tau) $ as defined in the text.
Indirect upper limits on the mixing angles from flavour-diagonal
observables, obtained in the models used, are taken into account. 
}}
\end{figure}
%=========================================================================
The new non-diagonal terms arise in diagrams D1 and D3 of Figure \ref{fig1}:
\beq
{\rm D1:} \ v_W(i,j) & = & 
-{\bf C}_{ij}\ \left[ \lambda_Q\ (C_{0}+C_{11}+C_{12}+C_{23})-2C_{24}+1\right] 
\nonumber \\
                    &   & +{\bf C}^*_{ij}\ \sqrt{\lambda_i\lambda_j}\ C_0,
\\ 
{\rm D3:} \ \ v_{\phi}(i,j)  & = &
-{\bf C}_{ij}\ \displaystyle\frac{\lambda_i\lambda_j}{2}\ C_0 
%\nonumber\\   & & 
+ {\bf C}^*_{ij}\displaystyle\frac{\sqrt{\lambda_i\lambda_j}}{2}
 \ \left[\lambda_Q \ C_{23}-2C_{24}+\displaystyle\frac{1}{2}\right].
\eeq
The matrices ${\bf B}$ and ${\bf C}$ are introduced in Appendix
\ref{app-genmaj}. 
Both contributions are quite similar to (\ref{D1}) and (\ref{D3}) when there
the right-handed couplings are retained.
The other contributions are the same as in the ordinary Dirac case.
The vertex reads:
\beq
{\cal V}_M(Q^2) &=&
\sum^{n_G+n_R}_{i,j=1}{\bf B}_{l_1 i} {\bf B}^*_{l_2 j}\
\left[ \delta_{ij}F(\lambda_i)+{\bf C}_{ij} G(\lambda_i,\lambda_j)
+{\bf C}^*_{ij} \sqrt{\lambda_i\lambda_j} H(\lambda_i,\lambda_j) \right],
\eeq
to be compared to the case of $n_H$ heavy sequential Dirac neutrinos 
(${\bf B}={\bf V}$):
\ba
{\cal V}(Q^2) &=&
\sum^{n_G+n_H}_{i=1}{\bf B}_{l_1 i} {\bf B}^*_{l_2 i}\ V(\lambda_i).
\ea

We now have to go into a specific model in order to make definite
predictions.
For the case chosen, namely the Standard Model extended by two heavy
Majorana singlets, the sums over virtual neutrinos  
involve $n_G$=3 light and $n_R$=2 heavy neutrinos.
The model is described in Appendix \ref{app-genmaj}.
For some important properties of the mixing matrices {\bf B} and
{\bf C} see  (\ref{p1})--(\ref{p4}).
%---------------------
The light neutrino sector is practically massless.
Using the properties (\ref{p1})--(\ref{p4}), one can then write the
form factor ${\cal V}_M$ in terms of the heavy-neutrino sector only.
Actually, it is:
\beq
\label{vhuge}
V(\lambda_i)=F(\lambda_i) + G(\lambda_i,\lambda_i).
\eeq
Further, the functions $F,G,H$ are uniquely defined and from
(\ref{p1})--(\ref{p4}), and taking $\lambda_i=0$ for $i=1,\ldots,n_G$, 
it is straightforward to prove that:
\ba
\label{branchM}
{\rm BR}(Z\to l^\mp_1 l^\pm_2) &=& \frac{\alpha^3_W
M_Z}{192\pi^2c^2_W\Gamma_Z}
\Big|\sum^{n_R}_{i,j=1}{\bf B}_{l_1 N_i} {\bf B}^*_{l_2 N_j} \nonumber\\
&&\times\big\{\delta_{N_i N_j}\ [F(\lambda_{N_i})-F(0)+G(\lambda_{N_i},0)+
G(0,\lambda_{N_i})-2G(0,0)]  \nonumber\\
&&\ \ +{\bf C}_{N_i N_j}\
[G(\lambda_{N_i},\lambda_{N_j})-G(\lambda_{N_i},0)
-G(0,\lambda_{N_j})+G(0,0)] \nonumber\\
&&\ \ +{\bf C}^*_{N_i N_j}\ \sqrt{\lambda_{N_i}\lambda_{N_j}} \
H(\lambda_{N_i},\lambda_{N_j})\big\}\Big|^2.
\ea

%h
%\ba
%\displaystyle\sum^{n_G+n_H}_{i=1} {\bf B}_{l_1 i} {\bf B}^*_{l_2 i}\ 
%here
%V(\lambda_i) 
%&=&
%\displaystyle\sum^{n_H}_{i=1} {\bf B}_{l_1 N_i} {\bf B}^*_{l_2 N_i}\ 
%\left[ V(\lambda_{N_i}) - V(0) \right], 
%\nl
%\\
%\displaystyle\sum^{n_G+n_R}_{i,j=1} {\bf B}_{l_1 i} {\bf B}^*_{l_2 i} 
%{\bf C}_{ij}\ 
%G(\lambda_i,\lambda_j) 
%&=&
% \displaystyle\sum^{n_R}_{i,j=1} {\bf B}_{l_1 N_i} {\bf B}^*_{l_2 N_j} 
%\{
%{\bf C}_{N_i N_j} [ G(\lambda_{N_i},\lambda_{N_j})-G(\lambda_{N_i},0)
%\nl &&
%-~G(0,\lambda_{N_j})+G(0,0) ]
%\nl &&
%+~ \delta_{N_i N_j} [ G(\lambda_{N_i},0)+ G(0,\lambda_{N_i})
%-2 G(0,0)
%]
%\}
%\nl
%\\
%\displaystyle\sum^{n_G+n_R}_{i,j=1} {\bf B}_{l_1 i} {\bf B}^*_{l_2 i} 
%{\bf C}^*_{ij} \sqrt{\lambda_i\lambda_j}\ 
%H(\lambda_i,\lambda_j)
%&=&
% \displaystyle\sum^{n_R}_{i,j=1} {\bf B}_{l_1 N_i} {\bf B}^*_{l_2 N_j}
%{\bf C}^*_{N_i N_j}\ \sqrt{\lambda_{N_i}\lambda_{N_j}}\
%H(\lambda_{N_i},\lambda_{N_j}).
%\nl
%\ea
It is possible to express the couplings 
${\bf B, C}$ on the right-hand side 
by the mass ratio $r = m_{N_2}^2 / m_{N_1}^2$ plus the three light-heavy
mixings $s_{\nu_e}, s_{\nu_{\mu}}, s_{\nu_{\tau}}$. 
%from flavour-diagonal observables.
The relations are explicitly given in
(\ref{bln1})--(\ref{cn2n2}). 
The upper limits for these mixings (given in Appendix \ref{app-b2}) 
have been used to obtain the graphs in Figure
\ref{figure-bergmt}(b).
The mass ratio $r$ has been taken as a free parameter.
Perturbative unitarity constrains the masses of the neutrinos so that they 
cannot be arbitrarily heavy.
%We have to take into account as an additional constraint that the
%heavy neutrinos may not be arbitrarily heavy.
This is discussed in Appendix \ref{app-puc}.
We see again from the figure that Giga--$Z$ has a discovery
potential, preferentially in the large neutrino mass
region. Similar curves can be obtained for $Z\to e\mu$ and
$Z\to e\tau$. 

The distinguished case of two Majorana singlet neutrinos with {\em equal}
masses, forming effectively a singlet Dirac particle \cite{Ilakovac:1995kj}, 
results for $r = 1$, see the solid line.
This line has been taken over from Figure \ref{figure-bergmt}(a), there
as the dashed line, in order to show that due to the different
coupling structure the simple sequential Dirac neutrino case does not
constitute a limiting case for large masses.
The deviations are due to the terms from the non-diagonal elementary
$Z$ couplings proportional to ${\bf C}$ and ${\bf C}^*$ in (\ref{branchM}). 
The ${\bf C}^*$ terms drop out for $r=1$.

In contrast, predictions for Majorana and Dirac 
neutrinos approach each other in the limit of small masses.
This may be seen from (\ref{branchM}) using the unitarity relations 
(\ref{p1})--(\ref{p4}) and the Taylor series expansion of the vertex function
in powers of the neutrino mass (Appendix \ref{sec-marie}).
This phenomenon is just another example of what is called in the
literature the ``practical Dirac-Majorana confusion theorem'' 
\cite{Kayser:1982br} (see also the recent discussion in
\cite{Zralek:1997sa,Czakon:1999cd} and references therein). 

At the end of this section, we would like to comment on the
literature for the reaction (\ref{eq1}).  
The early papers did not include the more ``realistic'' and physically most 
interesting models with Majorana masses.
These cases were studied in detail by \cite{Ilakovac:1995kj} and, 
in the context of left-right symmetric models, by \cite{Pilaftsis:1995tf}.
%In PRD 52 (1995), Eqn. (B1) a general expression is given for the process.
%We control all but the last 2 lines (relevant for LR models not
%considered by us).
%There, three times $C_{22}$ appears.
%First 2 should be $-C_{12}$, third should be absent.
%This is also controlled by the Dirac limit.
%Further, his explanation what are his $C$ functions is wrong by one
%sign for all but $C_{24}$.
While we reproduce the large mass limit given there, we obtain slight 
deviations in the medium mass case as given in Eqn. (B1) of
\cite{Pilaftsis:1995tf}. 
Further, 
%NPB 437 (1995) 
the Figures 7 and 8  in \cite{Ilakovac:1995kj} are not exactly reproducible in
the intermediate mass range. 
However, the large mass limit seems to be the only
potentially relevant case. 

%=======================================================================
\section{\label{sec-final}Summary}
\setcounter{equation}{0}
%=======================================================================
{From} our study
of the decays $Z \to e\mu, e\tau, \mu\tau$, in the context of the 
Giga--$Z$ option of the Tesla linear collider project, we conclude:

%All the cases discussed are model-dependent:
\begin{itemize}
\item
Neglecting the influence of the mixing angles, the expected branching
ratios depend on the neutrino masses $m_i$ and are of the order 
${\rm BR} \sim 10^{-5} (m_i/M_W)^4$ both in the small and large mass limits.
\item
If there exist only the known three generations of ultra-light
neutrinos, $m_{\nu} \ll M_W$, there is absolutely no hope to see any
effect. 
\item
For heavier neutrinos with masses of the order of the weak scale, 
$m_{\nu} \sim M_W$, 
one has to calculate the form factors describing the vertex without
any approximation. The necessary exact formulas have also been given.
\item
The light-heavy mixing angles are not so strongly restricted as with the 
minimal (one-family) seesaw mechanism when interfamily seesaw type models 
are considered.  But, unfortunately, these mixings have already been 
constrained to be very small, so that the lepton flavour-violating reactions 
under study are beyond the discovery reach of the Giga--$Z$ for 
$m_\nu\sim M_W$ .
\item
Since, ignoring the light-heavy mixings, the branching ratios are 
proportional to the fourth power of the neutrino masses, 
there is a discovery potential in the large mass case, $m_{\nu} \gg
M_W$. Nevertheless, they are limited in practice by potentially small mixing 
factors (bound independently of the heavy neutrino masses by other 
experiments) and upper neutrino mass bounds from unitarity considerations.
\item
In fact, there is an interplay between heavy masses and light-heavy mixings:
the mixings must be small for very large neutrino masses, since
otherwise the scattering matrix elements would grow above the unitarity limit.
%The mixing-matrix elements may modify the mass dependence of the
%vertex function by their own mass dependences. 
\end{itemize}
Summarizing, the Giga--$Z$ offers nearly three orders of magnitude 
gain of sensitivity compared to LEP~1. This opens quite
interesting opportunities to search for lepton-number changing processes,
if there exist heavy neutrinos sufficiently mixing with the light sector, 
within a quite broad allowed region according to present limits.

%=======================================================================
\section*{Acknowledgements}
We would like to thank 
J. Gluza, A. Pilaftsis, R. R\"uckl, and G. Wilson for helpful discussions. 
% A. Pilaftsis, 
% G. Wilson,
% J. Gluza,
% R. R\"uckl
We also wish to acknowledge J. Gluza and G. Wilson for carefully reading
the manuscript.
The work of J.I. has been partially supported by the Spanish 
  CICYT and Junta de Andaluc{\'\i}a, under contracts AEN96-1672 
  and FQM101, respectively.

%=======================================================================
\appendix
\def\theequation{\Alph{section}.\arabic{equation}}
%=======================================================================
\section{\label{app-feynman}
Lagrangians and Feynman Rules}
\setcounter{equation}{0}
%============================================================================
We make extensive use of the notion of Majorana particles,
since in general neutrinos may be of this type, and in fact in GUTs
often exactly this happens.
Majorana particles \cite{Majorana:1937vz} are neutral fermions $\psi$, 
fulfilling
\ba
\psi^c = \psi,
\label{selfc}
\ea
where $\psi^c$ is the charge-conjugate of $\psi$.

Some introduction on notations are given in Appendix \ref{dirac}.
We observe experimental evidences for ultra-light neutrinos in the
known fermion families on the one hand, and on the other there are
unifying theories with potentially ultra-heavy Majorana neutrinos. 
That both phenomena might be related will be made plausible by a toy
example in Appendix \ref{seesaw}.
A more general ansatz for Majorana mass terms is finally given in
Appendix \ref{app-genmaj}.  
%=======================================================================
\subsection{\label{dirac}Dirac neutrinos rewritten}
%=======================================================================
The mass-term Lagrangian for Dirac neutrinos is:
\ba
\label{freeD}
- {\cal L}_D &=& 
m_D \left( \overline{\nu_L} \nu_R +   \overline{\nu_R} \nu_L\right)
\equiv 
\frac{1}{2} \left(\overline{\chi^0}\right) 
\begin{pmatrix}
0      &  m_D
\\ 
m_D   & 0  
\end{pmatrix}
\left({\chi^0}\right),
\label{diracmassterm}
\ea
where we introduce (self-conjugate) Majorana fields $(\chi^0)$ 
\ba
\left({\chi^0}\right)
&=&
\begin{pmatrix}
\nu_L + \nu_L^c
\\
\nu_R + \nu_R^c
\end{pmatrix},
\ea
with 
\ba
\nu^c\equiv C\bar\nu^T
\ea
and $C$ being the charge-conjugation matrix.
The mass matrix can be brought into a diagonal form in the basis of $(\chi)$:
\ba
- {\cal L}_D  &=&
\frac{1}{2} \left(\overline{\chi}\right) 
\begin{pmatrix}
-m_D & 0
\\ 
0 & m_D
\end{pmatrix}
\left({\chi}\right) = \frac{1}{2} m_D \left(-{\bar \chi}_1 \chi_1 
+ {\bar \chi}_2 \chi_2 \right) 
\nl &=&
\frac{1}{2} \left(\overline{\xi}\right) 
\begin{pmatrix}
m_D & 0
\\ 
0 & m_D
\end{pmatrix}
\left({\xi}\right) = \frac{1}{2} m_D \left({\bar \xi}_1 \xi_1 
+ {\bar \xi}_2 \xi_2 \right).
\ea
In the last step, the field component $\xi_1$ is introduced as a
chiral transform of $\chi_1$  
in order to make the positive mass eigenvalue explicit.
The fields $(\chi)$ and $(\chi^0)$ are related by the unitary matrix ${\bf U}$, 
and $\xi_1$ additionally by $\gamma_5$: 
\ba
(\chi^0)={\bf U}\ (\chi);\quad 
{\bf U} &=& 
\frac{1}{\sqrt{2}}
\left(\begin{array}{rr}
1 & 1
\\ 
-1 & 1
\end{array}\right),
\ea
\ba
\xi_1 = \gamma_5 \chi_1 &=&
\frac{1}{\sqrt{2}}\left(-\nu_L+\nu_L^c-\nu_R+\nu_R^c\right)  , 
\\
\xi_2 = \chi_2 &=&
\frac{1}{\sqrt{2}}\left(\nu_L+\nu_L^c+\nu_R+\nu_R^c\right).
\label{chiraltransf}
\ea
With the above chain of equations we have shown that one ordinary
Dirac neutrino is equivalent to two Majorana neutrinos of equal mass
and opposite CP parities. 
Evidently, if the mass matrix is not of type (\ref{diracmassterm}),
then true Majorana particles are realized.

%=======================================================================
\subsection{\label{seesaw}The seesaw mechanism}
%=======================================================================
The seesaw mechanism \cite{Yanagida:1980xy,Gell-Mann:1980vs,Mohapatra:1980ia}
allows to understand the lightness of the known neutrinos by the
introduction of heavier ones.  
However, it is not the only possible solution to this puzzle; see
e.g. \cite{Tommasini:1995ii,Roulet:1999aa} for a nice discussion. 

In Appendix \ref{dirac} it was shown that a Dirac-neutrino mass term may  
be written as symmetric 2$\times$2 matrix with only off-diagonal elements.
Consider now the case where only {\em one} right-handed singlet
neutrino is added to the ordinary left-handed doublets of the SM.
Let's take one generation of left-handed light neutrinos for 
simplicity and with no loss of generality.  
The corresponding mass matrix is in general
%(\ref{massm}) 
\beq
{\bf M} = \left(\bea{cc} m_{L} & m_{D} \\ 
                         m_{D} & m_{R}    \eea\right) 
\eeq
and can be diagonalized by
\ba
{\bf U}&=&\left(\bea{rr} \cos\theta_\nu & \sin\theta_\nu \\
                        -\sin\theta_\nu &  \cos\theta_\nu \eea\right),
\label{diagU}
\ea
with
\ba
\tan 2\theta_\nu &=& \frac{2 m_{D}}
{m_R-m_L},\quad \cos2\theta=\frac{m_R-m_L}{\sqrt{(m_R-m_L)^2+4m^2_D}}
\label{tantwo}
\ea
yielding two eigenstates with different masses
\beq
m_\nu,m_N &=&\frac{1}{2}\left\{m_{R}+m_{L} \mp \left[\left(m_{R}-m_{L}
\right)^2+4m^2_{D}\right]^{1/2}\right\}.
\label{meigen}
\eeq
That is,
the most general mass term of one four-component self-conjugate field
describes two Majorana particles with different masses.
There is a particular configuration that corresponds to one Dirac
neutrino, as shown in Appendix \ref{dirac}. 

In the SM, $m_{L}=0$ since the SM Higgs sector consists of a Higgs
doublet (see Appendix \ref{app-genmaj}). Take now $m_{R}\gg m_{D}$. 
Then, the two physical states\footnote{ 
Actually, one of the mass eigenvalues in (\ref{meigen}) is negative. This is 
not a problem since then the true  mass eigenstate is a chiral transform of 
the original field which has a mass term of opposite sign as in 
(\ref{chiraltransf}) \cite{Gribov:1969kq}.} 
are a light and a heavy neutrino with 
masses
\ba
m_\nu &\simeq& m^2_{D}/m_{R},
\\
m_N  &\simeq& m_{R},
\ea
and the light-heavy mixing angle is:
\ba
s_{\nu}\equiv\sin\theta_\nu \simeq m_{D}/m_{R} \simeq \sqrt{m_\nu/m_N}.
\ea 
This is the minimal {\em seesaw} mechanism: 
the larger the heavy scale is, the smaller the light neutrino masses are,
and, at the same time, the smaller the mixing becomes. 

Actually, by taking 
$m_N\gsim M_Z$ and the present bounds on the light neutrino masses
\cite{PDG:1998aa},  
\ba
m_{\nu_e} &\lsim& 2.5\   \mbox{eV}, 
\\
m_{\nu_\mu} &\lsim& 160\ \mbox{keV}, 
\\
m_{\nu_\tau} &\lsim& 15\ \mbox{MeV},
\label{lmassbound}
\ea
we get from (\ref{tantwo}) the light-heavy mixing angles: 
\ba
s^2_{\nu_e}&\lsim& 3\times10^{-11},
\\
s^2_{\nu_\mu}&\lsim& 2\times10^{-6},
\\
s^2_{\nu_\tau}&\lsim& 2\times10^{-4}.
\ea
These mixing angles are too small to be constrained by the
experimental LEP and low-energy limits  
(\ref{hlmixa})--(\ref{hlmixc}) \cite{Nardi:1992rg,Nardi:1994iv}.

Nevertheless, one might have $m_{L}\ne 0$, by introducing a Higgs triplet
in the SM,~\footnote{The introduction of a Higgs triplet affects the value 
of the parameter $\rho=M^2_W/(M^2_Zc^2_W)$.} and hope for a conspiracy between 
$m_{L}$ and $m_R$ 
to get light-heavy mixings of order one. This does not seem very reasonable.
Thus, the minimal seesaw apparently lacks of phenomenological interest. 
%=======================================================================
\subsection{\label{app-genmaj}Majorana neutrinos}
%=======================================================================
In the {\em interfamily} seesaw-type models, several right-handed 
neutrinos are introduced and large light-heavy mixings can be obtained even 
for $m_{L}=0$ \cite{Buchmuller:1991tu,Pilaftsis:1992ug,Heusch:1994qu}.

Therefore following \cite{Ilakovac:1995kj}, we now introduce in our
study {\em two} 
heavy right-handed singlet neutrinos and treat the light-heavy mixings 
$s_{\nu_l}$ and the heavy masses $m_{N_1}, m_{N_2}$ as free phenomenological 
parameters.

Consider an extension of the SM that incorporates $n_R$ right-handed neutrinos 
$\nu_R$ (singlets under SU(2)$\otimes$U(1)) to the already present set of
$n_G=3$ generations of left-handed neutrino doublets. We keep the Higgs sector 
untouched.

It is convenient to arrange all the independent neutrino degrees of freedom
into two vectors of {\em left-handed} fields:
\ba
(\nu^0) &\equiv& (\nu_e,\nu_\mu,\nu_\tau),
\\
(N^0) &\equiv& (\nu^c_R)_i \quad i=1,\dots,n_R.
\ea
The charge-conjugates of right-handed neutrinos have been introduced:
\ba
\nu^c_R \equiv C\overline{\nu_R}^T.
\ea
The conjugate field has chirality and 
lepton/fermion numbers opposite to the original field (e.g. the $\nu^c_R$'s are
left-handed).

In the basis
\ba
(n^0) =(\nu^0,N^0)
\ea
the most general mass-term Lagrangian reads:
\beq
-{\cal L}_M&=&\frac{1}{2}(\overline{{n^0}^c})\ {\bf M}\ ({n^0})+\mbox{h.c.}
            = \frac{1}{2}(n^0)^T C\ {\bf M}\ (n^0)+\mbox{h.c.},
\eeq
where the Majorana fields
\ba
(\chi^0) \equiv (n^0)+({n^0}^c),
\ea 
containing both chiralities, have been introduced.

The mass matrix ${\bf M}$ is symmetric and can be written in a block form:
\beq
{\bf M}=\left(\bea{cc} {\bf m}_{L} & {\bf m}^T_{D} \\ 
                     {\bf m}_{D} & {\bf m}_{R} \eea\right).
\label{massm}
\eeq
It has dimension $n_G+n_R$ and can be diagonalized by a unitary matrix
${\bf U}$ of the same dimension,
\beq
\hat{\bf M} = {\bf U}^T {\bf M} {\bf U} = ({\bf U}^*)^{\dagger}{\bf M}{\bf U}
&=&{\bf diag}(m_{\nu_e},m_{\nu_\mu},m_{\nu_\tau},m_{N_1},m_{N_2})\nonumber\\
 &\simeq&{\bf diag}(0,0,0,m_{N_1},m_{N_2}),
\eeq
since we are mostly interested in a heavy neutrino sector consisting
of two right-handed neutrinos.
The interaction eigenstates, in terms of the 
left- and right-handed components of the
mass eigenstates, are given by: 
\beq
\label{Lh}
\mbox{Left-handed: } (n^0)&\equiv&\left(\bea{c} \nu^0\\ 
                N^0 \eea\right)=\left(\bea{c} \nu_L \\ 
                \nu^c_R \eea\right)
                ={\bf U}\ P_L\ (\chi)
\\
\label{Rh}
\mbox{Right-handed: } ({n^0}^c)&\equiv&\left(\bea{c} {\nu^0}^c \\ 
                {N^0}^c \eea\right)=\left(\bea{c} \nu^c_L \\ 
                \nu_R \eea\right)
                ={\bf U}^*\ P_R\ (\chi),
\eeq
with 
\ba
P_{R,L} \equiv \frac{1}{2}(1\pm\gamma_5).
\ea
The diagonal blocks in ${\bf M}$, connecting states with opposite fermion
number $({\bf m}_L,{\bf m}_R)$ are  called ``Majorana" mass terms, while the 
off-diagonal ones $({\bf m}_D)$ are ``Dirac" mass terms that conserve fermion 
number.\footnote{
Non-diagonal Dirac mass terms produce transitions between states with 
different individual lepton numbers but the total lepton number, i.e. the 
fermion number, is conserved.}
The mass terms arise, after the spontaneous symmetry breaking (SSB) of
SU(2)$_L\otimes$U(1)$_Y\to$U(1)$_Q$, from  
the Yukawa coupling of the fermion fields to the neutral Higgs field. 
The SU(2)$_L\otimes$U(1)$_Y$ quantum numbers of a Higgs doublet and the 
fermion bilinears that can be constructed from the doublet and singlet 
neutrinos are
\beq
\Phi=\left(\bea{c} \phi^+ \\
                   \phi^0 \eea\right) &\sim& ({\bf 2},1) \\
\Delta L=\ \ 0:\quad \overline{\nu_R}\nu_L &\sim& ({\bf 1},0)\otimes({\bf 2},-1)=
                ({\bf 2},-1) \\
\Delta L=\pm2:\quad \overline{\nu^c_L}\nu_L &\sim& 
({\bf 2},-1)\otimes({\bf 2},-1)=({\bf 1},-2)\oplus ({\bf 3},-2)\\
\Delta L=\pm2:\quad \overline{\nu^c_R}\nu_R & \sim& 
({\bf 1},0)\otimes({\bf 1},0)=({\bf 1},0) \quad (\mbox{neutral}).
\eeq
Therefore in a model with only Higgs doublets, the entry ${\bf m}_{L}={\bf 0}$. 

Our convention for the covariant derivative acting on a fermion doublet is
\beq
D_\mu=\partial_\mu+ig\frac{\vec\tau}{2}\cdot\vec{W}_\mu+ig'\frac{Y}{2}B_\mu
\eeq
with 
\ba
Q_f = I^{f_L}_3+Y/2. 
\ea
After SSB one gets 
\ba
e &=& g s_W = g'c_W,
\\
\left(\bea{c}Z_\mu\\A_\mu\eea\right)
&=&\left(\bea{rr} c_W & -s_W\\s_W & c_W\eea
\right)\left(\bea{c}W^3_\mu\\B_\mu\eea\right),
\\
W^\mp_\mu &=& \frac{1}{\sqrt{2}}\left(W^1_\mu \pm iW^2_\mu\right).
\ea
The interaction Lagrangians of neutrinos with $W$, $Z$ and charged Goldstone 
bosons $\phi^\pm$, in the weak basis\footnote{
The interaction of the $Z$ to two (right-handed) charge-conjugate neutrinos 
(non-existent in the ordinary SM) has been included in (\ref{ojo}), making use
of the bilinear transformation properties under charge-conjugation:
$\bar\psi\gamma^\mu P_{L,R}\psi \to - \overline{\psi^c}\gamma^\mu P_{R,L}
\psi^c$.}
\beq
{\cal L}_W &=& -\frac{g}{\sqrt{2}}W^-_\mu \sum^{n_G}_{i,j=1}
\bar{l}_i\gamma^\mu P_L \nu_{L_j} + \mbox{h.c.},
\\
{\cal L}_Z &=& -\frac{g}{2 c_W} Z_\mu \sum^{n_G}_{k=1}\ \left[
\overline{\nu_{L_k}}\gamma^\mu P_L\nu_{L_k} - 
{\overline{\nu^c_{L_k}}}\gamma^\mu P_R{\nu^c_{L_k}}
\right],
\label{ojo}
\eeq
become, in terms of physical neutrinos $\chi_j$ with masses $m_j$, 
$j=1,...,n_G+n_R$:
\beq
\label{lw}
{\cal L}_W &=& -\frac{g}{\sqrt{2}}W^-_\mu \sum^{n_G}_{i=1}\sum^{n_G+n_R}_{j=1}
{\bf B}_{l_i j}\bar{l}_i\gamma^\mu P_L\chi_j + \mbox{h.c.},
\\
\label{lz}
{\cal L}_Z &=& -\frac{g}{2 c_W} Z_\mu \sum^{n_G+n_R}_{i,j=1}\ 
\bar\chi_i\gamma^\mu(P_L{\bf C}_{ij} - P_R {\bf C}^*_{ij})\chi_j,
\\
\label{lh}
{\cal L}_{\phi^\pm} &=&-\frac{g}{\sqrt{2}}\phi^-\sum^{n_G}_{i=1}
\sum^{n_G+n_R}_{j=1}
{\bf B}_{l_i j}\bar{l}_i\left(\frac{m_{l_i}}{M_W} P_L -\frac{m_j}{M_W} P_R 
\right)\chi_j + \mbox{h.c.},
\eeq
with 
\beq
\label{blj}
{\bf B}_{l_i j} &\equiv&\sum^{n_G}_{k=1} {\bf V}^*_{l_i k} {\bf U}_{kj} ,
\\
\label{cij}
{\bf C}_{ij} &\equiv& \sum^{n_G}_{k=1} {\bf U}^*_{ki} {\bf U}_{kj} .
\eeq
${\bf V}_{l_i j}$ and ${\bf B}_{l_i j}$ are the leptonic CKM mixing matrices
and its {\em generalized} version, respectively
\cite{Schechter:1980gr,Pilaftsis:1992ug}.  
For Dirac particles ${\bf B}={\bf V}$, ${\bf C}_{ij}=
\delta_{ij}$ and ${\bf C}^*_{ij}=0$. For Majorana particles, in contrast, there 
are NC couplings of different flavours (FCNC) and with both left- and 
right-handed components.

The matrix ${\bf V}_{l_i j}$ is quadratic of dimension $n_G$ and 
${\bf B}_{l_i j}$ is rectangular $n_G\times (n_G+n_R)$ 
and incorporates lepton-flavour changing mixings. 
The matrix ${\bf C}_{ij}$
is quadratic, has dimension $(n_G+n_R)$, and causes flavour non-diagonal
$Z\chi_i\chi_j$ interactions. 

These interaction Lagrangians involve Majorana fermions that 
complicate the evaluation of $S$ matrix elements, since extra Wick 
contractions survive in comparison to the case with only Dirac fermions.
Following \cite{Denner:1992me,Denner:1992vz}, one can write 
Feynman rules resembling the ones of the Dirac fermions, based on the
well defined {\em fermion flow}, rather than on the {\em fermion number 
flow} which is not preserved in the vertices with Majorana fermions:
after fixing an arbitrary orientation (fermion flow) for a given diagram, the
vertices can be read off from the Lagrangian as usual, {\em but} for every
vertex $\bar{f}_1\Gamma f_2$ one has to add the reversed one, 
$\overline{f^c_1}\Gamma' f^c_2$, with $\Gamma'=C\Gamma^T C^{-1}$.
The fermion propagators are the usual ones.
This effectively yields the same result for the vertex $Wl_i\chi_j$ but a 
{\em factor two} larger for the vertex $Z\chi_i\chi_j$ in comparison to the 
case of Dirac neutrinos, since for two Majorana fermions ($\chi=\chi^c$)
$\Gamma=\Gamma'$.

{\em Important note}: elsewhere in the text we refer to $\nu_i$ and $N_i$
as the neutrino physical states, rather than $\chi_i$, to simplify
the presentation, but with no possible confusion, since we always work
in the physical basis.

The matrices ${\bf B}$ and ${\bf C}$ obey a number of useful relations
\cite{Korner:1993an}:
\ba
\sum^{n_G+n_R}_{j=1} {\bf B}_{l_1 j} {\bf B}^*_{l_2 j} &=& \delta_{l_1 l_2} ,
\label{p1}
\\
\sum^{n_G+n_R}_{k=1} {\bf C}_{ik} {\bf C}^*_{jk} &=& 
\sum^{n_G}_{k=1} {\bf B}_{l_k i} {\bf B}^*_{l_k j}\ =\ {\bf C}_{ij} ,
\label{p2}
\ea
\ba
\sum^{n_G+n_R}_{k=1} {\bf B}_{l k} {\bf C}_{kj} &=& {\bf B}_{l j} ,
\label{p3}
\\
\sum^{n_G+n_R}_{k=1} m_k {\bf C}_{ik} {\bf C}_{jk} &=&
\sum^{n_G+n_R}_{k=1} m_k {\bf B}_{l k} {\bf C}^*_{ki}\ =\
\sum^{n_G+n_R}_{k=1} m_k {\bf B}_{l_1 k} {\bf B}_{l_2 k}\ =\ 0.
\label{p4}
\ea

In the case of $n_R=2$ the matrix elements involving heavy neutrinos can
be obtained from (\ref{p2}), (\ref{p4}) in terms of the light-heavy mixing 
angles (\ref{lihe}) and the ratio of the two heavy masses squared 
$r\equiv m^2_{N_2}/m^2_{N_1}$, assuming that the light sector consists of
massless neutrinos \cite{Ilakovac:1995kj}:
\ba
\label{bln1}
{\bf B}_{l_k N_1} &=& \frac{r^{1/4}}{\sqrt{1+r^{1/2}}}\
s_{\nu_k}, 
\\
\label{bln2}
{\bf B}_{l_k N_2} &=& \frac{i}{\sqrt{1+r^{1/2}}}\ s_{\nu_k} ,
\ea
and
\ba
{\bf C}_{N_1 N_1} &=& \frac{r^{1/2}}{1+r^{1/2}}\sum^{n_G}_{k=1}
                                                         s^2_{\nu_k} ,
\label{cn1n1} 
\\
{\bf C}_{N_2 N_2} &=& \frac{1}{1+r^{1/2}}\sum^{n_G}_{k=1}
                                                         s^2_{\nu_k} ,
\\
\label{cn2n2} 
{\bf C}_{N_1 N_2} &=& -{\bf C}_{N_2 N_1}\ =\ \frac{ir^{1/4}}{1+r^{1/2}}
                                        \sum^{n_G}_{k=1} s^2_{\nu_k}.
\ea

%=======================================================================
\section{\label{app-exp}Constraints on Heavy Neutrinos}
\setcounter{equation}{0}

\subsection{\label{app-b2}Experimental bounds}

Several kinds of constraints on heavy-neutrino masses and light-heavy
mixings can be obtained from experiment.

Direct production searches establish the following 
limits on the neutrino masses at 95\% c.l. \cite{PDG:1998aa}:
\ba
\mbox{Stable neutrinos: }m_N&>&45.0\ (39.5)\  \mbox{GeV}\\
\mbox{Unstable neutrinos: }m_N&>&69.0\ (58.2)\  \mbox{GeV},
\ea
for Dirac (Majorana) particles, respectively.

A general formalism to describe light-heavy mixings was developed in 
\cite{Langacker:1988ur,Langacker:1988up}. The mixing angles 
in our notation \cite{Ilakovac:1995kj} correspond to
\beq
s^2_{\nu_k} \equiv \sum_{i} |{\bf B}_{l_k N_i}|^2.
\label{lihe}
\eeq
Indirect constraints on the masses and bounds on the mixings are provided by 
two categories of LEP and low energy experiments:

(i) Flavour-diagonal processes. They include mass-independent and
model-in\-de\-pen\-dent light-heavy mixing constraints \cite{Langacker:1988ur,
Nardi:1992rg,Nardi:1994iv,Bergmann:1998rg} from tests of lepton 
universality and CKM unitarity and measurements of the $Z$ boson invisible 
width, as well as other less sensitive studies like $W$ mass measurements 
and low energy experiments like neutrino scattering, atomic parity violation, 
etc. Flavour-conserving leptonic decays $Z\to l^-l^+$ depend on masses
and mixings through loop contributions and provide alternative constraints
\cite{Bhattacharya:1995bj}.

(ii) Flavour-changing processes. They include rare processes
like $\mu\to e\gamma$, $\mu\to ee^+e^-$, $\mu\to e$ conversion
in nuclei, $\tau\to l_a l^+_b l^-_c$ and $Z\to l^-_a l^+_b$ 
\cite{Langacker:1988ur,
Tommasini:1995ii}. They are mass-dependent (except for $\mu\to e\gamma$ to a 
good approximation). One can get also less stringent light-heavy mixing 
constraints from oscillation experiments \cite{Langacker:1988up,Bergmann:1998rg}.

The most stringent {\em present} bounds on the light-heavy mixings are provided 
by the flavour-diagonal processes. Exceptions are the ones  
involving the first two lepton families such as $\mu\to e\gamma$, $Z\to e\mu$ 
and $\mu\to ee^+e^-$ \cite{Tommasini:1995ii}.

For illustration we show an example of the flavour-diagonal constraints.
The effective muon decay constant $G_{\mu}$ is related to the coupling
$G_F$ of the standard model by
\cite{Langacker:1988ur,Nardi:1992rg,Nardi:1994iv}: 
\ba
G_{\mu} = c_{\nu_e} c_{\nu_{\mu}} G_F.
\ea
The unitarity constraint for the first row of the CKM quark-mixing matrix 
implies \cite{Nardi:1994iv}
\ba
\sum_{i=1}^3|V_{ui}|^2 &=& \left(\frac{c_{\nu_e}G_F}{G_{\mu}}\right)^2
= \frac{1}{c^2_{\nu_{\mu}}}
= 0.9992 \pm 0.0014.
\ea

As a summary, from \cite{Langacker:1988ur,Nardi:1992rg,Nardi:1994iv}
(90\% c.l.):
\beq
s^2_{\nu_e}  &<& 0.0071\ (0.005),
\label{hlmixa}
\\
s^2_{\nu_\mu}&<& 0.0014,
\label{hlmixb}
\\
s^2_{\nu_\tau}&<& 0.033\ (0.01),
\label{hlmixc}
\eeq
where the most conservative bounds are obtained assuming any kind of
heavy neutrinos and the ones in brackets correspond to the case of SU(2) 
singlets.

From the most recent update\footnote{
These latest bounds are more conservative than the earlier ones in
\cite{Nardi:1994iv} due to the fact that present determinations of the 
elements of the first row of the CKM matrix are not compatible with
unitarity, and hence this constraint is eliminated from the analysis.}
by \cite{Bergmann:1998rg}, assuming only 
heavy singlets, one gets (90\% c.l.):
\beq
\label{b11}
s^2_{\nu_e}&<&0.012,
\label{hlmix2a}
\\
\label{b12}
s^2_{\nu_\mu}&<&0.0096,
\label{hlmix2b}
\\
\label{b13}
s^2_{\nu_\tau}&<&0.016.
\label{hlmix2c}
\eeq

A final remark is in order here. Using the Schwartz inequalities 
\cite{Langacker:1988up},
\beq
|\sum_{i} {\bf B}_{l_a N_i}{\bf B}^*_{l_b N_i}|^2 < 
s^2_{\nu_a} s^2_{\nu_b},
\label{fd2}
\eeq
one can infer {\em indirect} upper limits on the off-diagonal mixings 
(relevant for the flavour-changing processes) from the previous 
flavour-diagonal constraints.
Nevertheless, for our scenario (iii), as already mentioned, the mixings can
be obtained exactly from the properties of ${\bf B}$ and ${\bf C}$ and
such inequalities are not needed.

%==========================================================================
\subsection{\label{app-puc}Decoupling and neutrino-mass upper limits from
perturbative unitarity} 
%--------------------------------------------------

The heavy-neutrino masses are restricted by
the perturbative unitarity condition on the decay width of heavy 
neutrinos \cite{Chanowitz:1979mv,Durand:1990zs,Durand:1992wb,Fajfer:1998px,
Ilakovac:1999md},
\beq
\Gamma_{N_i}\leq \frac{1}{2} m_{N_i}.
\label{pub}
\eeq
The total decay width of a heavy Dirac neutrino (four d.o.f.) with 
mass $m_{N_i}\gg M_W,M_Z,M_H$ is \cite{Fajfer:1998px,Ilakovac:1999md,
Buchmuller:1991vh,Pilaftsis:1992ug}:
\beq
\Gamma_{N_i}&=&\sum^{3}_{k=1}\Gamma(N_i\to l^-_k W^+)
+\sum^{3}_{k=1} \left[ \Gamma(N_i\to \nu_k Z)
                            +\Gamma(N_i\to \nu_k H)\right] 
\nonumber \\
 &\simeq&\frac{\alpha_W}{8M^2_W}m^3_{N_i}\sum^{3}_{k=1}|{\bf B}_{l_k N_i}|^2,
\eeq
and a factor two larger for a heavy Majorana neutrino (two d.o.f.).
Therefore, the perturbative unitarity bound expressed in
(\ref{pub}) reads 
\beq
m^2_{N_i}\sum^{3}_{k=1}|{\bf B}_{l_k N_i}|^2=m^2_{N_i}{\bf C}_{N_i N_i}
\leq 
\left\{
\begin{array}{l} 4M^2_W/\alpha_W \mbox{ for a Dirac neutrino}\\ \\
                 2M^2_W /\alpha_W \mbox{ for a Majorana neutrino},
\end{array} 
\right.
\label{pub2}
\eeq               
(no summation over repeated indices is understood) which shows implicitly that 
{\em heavy neutrinos
decouple} \cite{Fajfer:1998px,Ilakovac:1999md}, in accordance with the 
Appelquist-Carazzone theorem \cite{Appelquist:1975tg,Senjanovic:1980yq}: the 
unacceptable large-mass behaviour of the amplitudes
($\sim m^2_N$) is actually cured when the light-heavy mixing ($\sim m^{-2}_N$)
is taken into account \cite{delAguila:1982yu,Cheng:1991dy}.

Taking the values of ${\bf B}_{l_k N_i}$
in terms of the light-heavy mixing angles (\ref{lihe}) for one Dirac or
for two heavy Majorana neutrinos (\ref{bln1}), (\ref{bln2}) one can get the 
following upper limits:
\beq
m^2_{N}&\leq&\frac{4 M^2_W}{\alpha_W}
\left[\sum^{3}_{k=1} s^2_{\nu_k}\right]^{-1}
\label{pubdi}
\eeq
for a heavy Dirac neutrino, and
\beq
m^2_{N_1}\equiv\frac{1}{r}m^2_{N_2}&\leq&\frac{2 M^2_W}{\alpha_W}
\frac{1+r^{1/2}}{r}
\left[\sum^{3}_{k=1} s^2_{\nu_k}\right]^{-1}
\label{pubma}
\eeq
for two heavy Majorana singlets. The latter bound is very stringent
when $m_{N_1}$ and $m_{N_2}$ are very different.
The upper mass limits on heavy neutrinos, from (\ref{pubdi}), (\ref{pubma}) 
are then:
\beq
 \mbox{scenario (ii): }~~~~~~~m^2_N     &\lsim& (4.2,4.4\ \mbox{TeV})^2 \\
\mbox{scenario (iii): }m^2_{N_1}\equiv\frac{1}{r} m^2_{N_2} 
&\lsim& \frac{1+r^{1/2}}{r}\times (3.0,3.1\ \mbox{TeV})^2,
\eeq
using the bounds (\ref{hlmixa})--(\ref{hlmixc}) and 
(\ref{hlmix2a})--(\ref{hlmix2c}),  
respectively.
%=======================================================================
\section{\label{app-vertex}The Vertex Function}
\setcounter{equation}{0}
%=======================================================================
%\subsection{\label{sec-tensor}The tensor integrals}
We use the notations of \cite{Denner:1993kt}.
The vertex function contains the following one-loop integrals in $D$
dimensions \cite{'tHooft:1972fi,Passarino:1979jh}:
\beq
\frac{i}{16\pi^2}A(m^2_0)&=&
\mu^{4-D}\int\frac{d^D q}{(2\pi)^D}
\frac{1}{{\cal D}_0},  
\\
\frac{i}{16\pi^2}\{B_0,B^\mu\}(p_1^2;m^2_0,m^2_1)&=&
\mu^{4-D}\int\frac{d^D q}{(2\pi)^D}
\frac{\{1,q^\mu\}}{{\cal D}_0 {\cal D}_1},  
\\
\frac{i}{16\pi^2}\{C_0,C^\mu,C^{\mu\nu}\}
(p_1^2,Q^2,p^2_2;m^2_0,m^2_1,m^2_2)&=&
\mu^{4-D}\int\frac{d^D q}{(2\pi)^D}
\frac{\{1,q^\mu,q^\mu q^\nu\}}{{\cal D}_0 {\cal D}_1 
{\cal D}_2},
\eeq
with 
\ba
\label{momenta}
Q^2=(p_2-p_1)^2
\ea
 and
\beq
{\cal D}_0&=&q^2-m^2_0+i\epsilon, 
\\
{\cal D}_1&=&(q+p_1)^2-m^2_1+i\epsilon, 
\\
{\cal D}_2&=&(q+p_2)^2-m^2_2+i\epsilon.
\eeq
They are decomposed into tensor integrals according to their Lorentz
structure:
\beq
B^\mu &=& p^\mu B_1,
\\
C^\mu &=& p^\mu_1 C_{11} + p^\mu_2 C_{12}, 
\\
C^{\mu\nu} &=& p^\mu_1 p^\nu_1 C_{21} + p^\mu_2 p^\nu_2 C_{22}
+ (p^\mu_1 p^\nu_2 + p^\mu_2 p^\nu_1) C_{23} + g^{\mu\nu} C_{24}.
\eeq
We employ the following dimensionless tensor integrals and
their  abbreviations:
\beq
B_1 &\equiv& B_1(0;\lambda_i,1)
\nl
&=& B_1(0;m^2_i,M^2_W),
\\
C_{\{0,11,12,21,22,23\}}   &\equiv& 
C_{\{...\}}(0,\lambda_Q,0;1,\lambda_i,\lambda_j) 
\nl
&=& M^2_W C_{\{...\}}(0,Q^2,0;M^2_W,m^2_i,m^2_j),
\\
C_{24}  &\equiv& 
C_{24}(0,\lambda_Q,0;1,\lambda_i,\lambda_j)
\nl
&=& C_{24}(0,Q^2,0;M^2_W,m^2_i,m^2_j),
\\
\bar C_{\{0,11,12,21,22,23\}}&\equiv& 
C_{\{...\}}(0,\lambda_Q,0;\lambda_i,1,1)
\nl
&=& M^2_W C_{\{...\}}(0,Q^2,0;m^2_i,M^2_W,M^2_W),
\\
\bar C_{24}  &\equiv& 
\bar C_{24}(0,\lambda_Q,0;\lambda_i,1,1)
\nl
&=&C_{24}(0,Q^2,0;m^2_i,M^2_W,M^2_W).
\eeq
On the $Z$ mass shell it is $\lambda_Q = \lambda_Z$.
The $\lambda_Q, \lambda_Z, \lambda_i$ 
%=M^2_Z/M^2_W$ 
%and $\lambda_i$ %=m^2_i/M^2_W$.
are introduced in (\ref{delta}), (\ref{deltaq}), and (\ref{deltaz}).

The explicit expressions for the loop functions are:
% in notations of Jose.
% no change in sign: A,B1,c24. all other change
% it from Minkowski to euklidian metrics
\ba
C_{\{0,11,23\}} &=& \int_0^1dx\int_0^x\frac{dy}{\cal D}\{-1,y,-(1-x)y\},
\\
C_{12}(\lambda_i,\lambda_j) &=&  C_{11}(\lambda_j,\lambda_i),
\\
\label{uv-c24}
C_{24} &=& -\frac{1}{2(D-4)} - \frac{1}{2}\int_0^1dx\int_0^x dy \ln
{\cal D},
\ea
with
\ba
{\cal D} &=& D_{ijk}
\nl &=& 
\lambda_Q xy + (\lambda_k -\lambda_j)x+ (-\lambda_Q+\lambda_i-
\lambda_k)y + \lambda_j -  i\epsilon,
\ea
and the correspondences are: for the non-abelian diagrams (with
elementary $ZWW$, $ZW\phi$, $Z\phi\phi$ vertices),
($i,j$) are virtual $W,\phi$ bosons and $k$ a neutrino, while for the
abelian diagrams $k$ is the $W,\phi$ boson and ($i,j$) are neutrinos.
On the $Z$ boson mass shell:
\ba
{\cal D} &\equiv& D_{ijW} = 
\lambda_Z xy + (1-\lambda_j)x+ [-\lambda_Z+(\lambda_i-1)]y + \lambda_j -
i\epsilon, 
\\
{\cal{\bar D}}  &\equiv& D_{WWi} =  
\lambda_Z xy - (1-\lambda_i)x+ [-\lambda_Z-(\lambda_i-1)]y +1 -
i\epsilon.
\ea
In the Dirac case, it is $\lambda_i=\lambda_j$.
Further,
\ba
\label{uv-b1}
B_1 &=& \frac{1}{D-4} + \int_0^1 xdx \ln[x+\lambda_i(1-x) - i\epsilon] 
\nl
&=&  \frac{1}{D-4} + \frac{\lambda_i}{2(1-\lambda_i)}
\left(1+\frac{\lambda_i\ln\lambda_i}{1-\lambda_i} \right) -\frac{1}{4}.
\ea
As may be seen,
the tensor integrals $B_1$, $C_{24}$ and $\bar C_{24}$ are
ultraviolet--divergent in $D$ dimensions.  
We mention that the functions are defined here with Minkowskian
metric, so that all the functions introduced, except for $B_0, B_1, C_{24},
{\bar C}_{24}$, have different sign from those used in e.g. \cite{Mann:1984dvt}
(Euclidean metric).
%Their poles are given by:
%\ba
%\label{bd-div}
%B_1 &=& \frac{1}{D-4} + B'_1,
%\label{uv-b1}
%\\
%\left[ C_{24},~{\bar C}_{24}\right] &=& -\frac{1}{2(D-4)} + \left[
%C'_{24},~{\bar C}'_{24}\right] , 
%\label{uv-c24}
%\ea
%where $B'_1,C'_{24},{\bar C'_{24}}$ are the finite parts.
In principle, the divergent parts depend on the regularization scheme
and could differ by a finite, universal term yet.
%-----------------------------------------------------------------------------
%\subsection{The cancellation of the UV-divergences}
%-----------------------------------------------------------------------------
Recalling the UV-behaviour of the divergent integrals $B_1$, $C_{24}$ and
$\bar C_{24}$ (see (\ref{uv-b1}) and (\ref{uv-c24})), we get
divergent, mass-dependent contributions from individual diagrams to
the vertex $V$.

For the Dirac case:
\ba
%-v_W(i) &\sim& -(v_i+a_i)\times \left[\frac{1}{D-4} + \mbox{finite} \right] 
%-\nonumber\\
%-       &    & -(v_i-a_i)\times \left[ \mbox{finite} \right] 
%-\\
%-v_{WW}(i) &\sim& (2I^L_3)\ 2c^2_W\times \left[\frac{3}{D-4} + \mbox{finite} 
%-\right] 
%-\\
v_\phi(i) &\sim& -(v_i+a_i)\times \left[ \mbox{finite} \right]  
%\nonumber\\
%       &    & 
-(v_i-a_i)\times \left[ \frac{1}{2}\frac{\lambda_i}{D-4} 
       + \mbox{finite} \right] ,
\\
v_{\phi\phi}(i) &\sim&-(2I^{i_L}_3)\ (1-2s^2_W)\times 
\left[-\frac{1}{2}\frac{\lambda_i}{D-4} + \mbox{finite} \right], 
\\
%-v_{W\phi}(i) &\sim&-(2I^L_3)\ 2s^2_W\times \left[ \mbox{finite} \right] 
%-\\
v_\Sigma(i) &\sim& \frac{1}{2}(v_i+a_i-4c^2_W a_i)
\left[\frac{2}{D-4} + \frac{\lambda_i}{D-4} + \mbox{finite} \right] .
\ea
The sum of the terms proportional to $\lambda_i/(D-4)$ vanishes.
The constant divergent terms (not shown here) sum up for individual vertices
$V(\lambda_i)$, but vanish due to the unitarity of the mixing matrix
for the complete vertex function (\ref{order-V}):
\ba
{\cal V} &\sim& (2I^{i_L}_3)\ 4c^2_W\ \frac{1}{D-4}\ \delta_{l_1 l_2} = 0.
\ea

For the Majorana case ($2I^{i_L}_3=1$, $v_i=a_i=1/2$):
\ba
%-v_W(i,j) &\sim& -{\bf C}_{ij} \times \left[\frac{1}{D-4} +
%-\mbox{finite} \right]  
%-\nonumber\\
%-       &    & + {\bf C}^*_{ij}\times \left[ \mbox{finite} \right] 
%-\\
%-v_{WW}(i,j) &\sim& \delta_{ij}\ 2c^2_W\times \left[\frac{3}{D-4} + 
%-\mbox{finite} \right] 
%-\\
v_\phi(i,j) &\sim& -{\bf C}_{ij} \times \left[ \mbox{finite} \right]  
%\nonumber\\
%       &    & 
+ {\bf C}^*_{ij}\times \left[
       \frac{1}{2}\frac{\sqrt{\lambda_i\lambda_j}}{D-4} 
       + \mbox{finite} \right] ,
\label{c27}
\\
v_{\phi\phi}(i,j) &\sim&-\delta_{ij}\ (1-2s^2_W)\times 
\left[-\frac{1}{2}\frac{\lambda_i}{D-4} + \mbox{finite} \right] ,
\label{c28}
\\
%-v_{W\phi}(i,j) &\sim&-\delta_{ij}\ 2s^2_W\times \left[ \mbox{finite} \right] 
%-\\
v_\Sigma(i,j) &\sim& \delta_{ij}\ \frac{1}{2}(-1+2s^2_W)
\left[\frac{2}{D-4} + \frac{\lambda_i}{D-4} + \mbox{finite} \right] .
\label{c29}
\ea
The mass-dependent divergent terms in (\ref{c28}) and (\ref{c29}) cancel
each other, and the one proportional to ${\bf C}^*_{ij}$ in (\ref{c27})
drops out due to the unitarity condition (\ref{p4}) when the sum over virtual
neutrinos is performed in (\ref{calvm}). The constant divergent terms vanish
again due to the unitarity relations (\ref{p1})--(\ref{p4}):
\ba
{\cal V}_M &\sim& 4c^2_W\ \frac{1}{D-4}\ \delta_{l_1 l_2} = 0
\ea
%---------------------------------------------------------------------------
\section{\label{app-limits}The Vertex Function for Large and Small
Neutrino Masses}
%---------------------------------------------------------------------------
\subsection{\label{app-bigm}The vertex for large neutrino
mass, $\lambda_i \gg 1$ 
}
%---------------------------------------------------------------------------
We now consider the vertex function in the limit of large Dirac-neutrino masses
$(\lambda_i=\lambda_j)$. 
The leading terms of the one-loop functions for $\lambda_Q=\lambda_Z$
are:\footnote{
The functions ${\bar C_{11}}$, ${\bar C_{12}}$, ${\bar C_{23}}$ do not
contribute to the large mass limit of the vertex, and we reproduce only their
first leading terms, of order $\ln\lambda/\lambda$. If these functions are of 
relevance for an application, one should determine also the terms of order 
$1/\lambda$.}
\ba
C_0 &=& -\frac{1}{\lambda_i} 
+\frac{\ln\lambda_i}{\lambda_i^2}
-\left(12+\dz\right)\frac{1}{12\lambda_i^2} +\ldots,
\\
C_{11}&=&C_{12}=\frac{1}{4\lambda_i}+\ldots,
\\
C_{23} &=&-\frac{1}{18\lambda_i} +\ldots,
\\
C_{24} &=&
-\frac{1}{2(D-4)}
- \frac{1}{4}\ln\lambda_i 
+\frac{1}{8} 
+ \left(-9+\dz\right)\frac{1}{36\lambda_i}  +\ldots,
\\
B_1 &=&
\frac{1}{(D-4)}
+ \frac{1}{2}\ln\lambda_i 
-\frac{3}{4} + \frac{\ln\lambda_i}{\lambda_i} - \frac{1}{2\lambda_i}
+\ldots,
\\
{\bar C_0} &=& -\frac{\ln\lambda_i}{\lambda_i} -\left[1-4a(y) \right]
\frac{1}{\lambda_i} +\ldots,
\\
{\bar C}_{11}&=&{\bar C}_{12}=\frac{1}{2}\frac{\ln\lambda_i}{\lambda_i}
+\ldots,\\
{\bar C}_{23}&=&-\frac{1}{6}\frac{\ln\lambda_i}{\lambda_i}+\ldots,
\\
{\bar C}_{24} &=& 
-\frac{1}{2(D-4)}
-\frac{1}{4}\ln\lambda_i
+\frac{3}{8}
+ \left(-6+\dz\right)\frac{\ln\lambda_i}{12\lambda_i}
\nl&&+~ \left[-30+5\dz+24(4-\dz)a(y) \right]\frac{1}{72\lambda_i}
+\ldots,
\ea
with 
\ba
y &=& \sqrt{1/\dz  - 1/4},
\label{y}
\\
 a(y) &=& y \arctan(1/2y).
\ea
The large Dirac-neutrino mass limit of the vertex function is:
\ba
\label{vbig}
V(\lambda_i)
&=&
I_3^{\nu_L}
\Biggl[
\lambda_i
+
\left(3-\frac{\dz}{6} (1-2s_W^2) \right) \ln\lambda_i
\nl
&&+~\frac{1}{18} \left(-66 - \dz + 96 s_W^2 +5 s_W^2\dz \right)
\\
%\nl 
&&+~\frac{1}{3}
% in Ann.Phys. we had 32/8s_W^2\dza(y) instead of 32/3.
% this was wrong, a typo. see VQ82, 25-11-82-7.
\left(-8+2\dz-32 s_W^2-4s_W^2\dz\right) a(y) 
%\nl 
%&&
%%-~ \frac{7}{36}  s_W^2 |Q_i|\dz             
+ \frac{8c_W^2}{D-4}
\Biggr]
+ {\cal O}\left(\frac{\ln\lambda_i}{\lambda_i}\right).
\nonumber
\ea
Numerically, this means for the value $\lambda_Z = 1.286$:
\ba
\label{bigva0}
V(\lambda_N) - V(0)
&=&
\frac{1}{2}
\left[
\lambda_4 + 2.88 \ln \lambda_4 - 4.47 
%- 0.05 |Q_{\nu}|
%\right]
%\nl &&
-
(2.52 + 2.11 \times i)
% - (0.23 + 0.41 \times i) |Q_{\nu}|   
\right].
\label{d40}
\ea
Here, we subtracted from the large mass limit of the vertex function
its value at zero mass (\ref{a0}) in order to obtain (\ref{eq3-6}).
%---------------------------------------------------------------------------
\subsubsection{\label{app-zbb}A relation to the $Zb{\bar b}$ vertex}
%---------------------------------------------------------------------------
The loop contributions to the vertex function $V(\lambda)$ are
gauge-invariant. 
We calculate them in the 't Hooft-Feynman gauge.
They describe not only the flavour-changing $Zf{\bar f'}$ vertex, but also
the mass dependent terms of the flavour-diagonal vertex.
There is one case where the effect is quite visible, namely the
$Zb{\bar b}$ vertex with virtual $t$ quark exchanges.
The exact one-loop expression for the $t$ quark mass dependent part
${\cal W}$ of the 
$Zb{\bar b}$ vertex correction was first given in Eqn. (22) of
\cite{Akhundov:1986fct} (form factor $\delta \kappa$, calculated in
the unitary gauge).
The large $t$ quark mass limit is (Eqn. (2.4.30) of \cite{Bardin:1999yd}):
\ba
\label{vt}
{\cal M}(Z\to {\bar b}b) 
&\sim& \epsilon^{\mu} {\bar u}
\left[
\gamma_{\mu} v_b - \gamma_{\mu}\gamma_5 a_b 
+ \gamma_{\mu} (1-\gamma_5){\cal W}(\lambda_t,\lambda_Z)
\right]u,
\\ 
\frac{{\cal W}(\lambda_t,\lambda_Z)}{a_b} &=&
\frac{\alpha}{\pi}\frac{1}{4s_W^2}|P_{tb}|^2 {V}_t,
\\
{V}_t &=&
\frac{1}{2}\left[\lambda_t + \left(\frac{8}{3}+\frac{\lambda_Z}{6}\right) 
\ln\lambda_t + {\cal O}(1) \right].
\ea
This agrees with the leading terms in (\ref{vbig}).
 
%=======================================================================
\subsection{\label{sec-marie}The vertex for small neutrino mass,
$\lambda_i \ll 1$} 
%=======================================================================
The limits of the $C$ functions for $\lambda_i=\lambda_j$ and $\lambda_Q=
\lambda_Z\gg\lambda_i$ are:
\ba
C_{0} &=&
-c_{0}
+ \frac{2}{\dz}\lambda_i\ln\lambda_i
-
\frac{2}{\dz(1+\dz)}(1+\dz+\ln\dz-i\pi)\lambda_i
+\ldots,
\\
\dz C_{11}  &=& \dz C_{12} 
\nl &=& 
~-c_0 +1 - \ln\dz +i\pi
-\lambda_i\ln\lambda_i
+\left[c_0+\frac{2}{1+\dz}(\ln\dz-i\pi)\right]\lambda_i
\nl && 
+~\ldots,
\\
\dz^2 C_{23} &=&
-(\dz+2) c_0 + \frac{\dz}{2} + 2 - 2 (\ln\dz-i\pi)
\nl && 
-~ \left[- 4 (c_0-1) - 2 \frac{2+\dz}{1+\dz} (\ln\dz-i\pi)
\right]\lambda_i
+\ldots,
\\
4\dz C_{24} &=&
- \frac{2\dz}{D-4}
-2(1+\dz) c_0 +3\dz +2 -(\dz+2) (\ln\dz-i\pi)
\nl && 
+ 4\left[c_0 -1 +\ln\dz-i\pi\right]\lambda_i
+\ldots,
%\\
\ea

\ba
B_{1} &=& 
\frac{1}{D-4}
-\frac{1}{4}  + \frac{1}{2}\lambda_i
+\ldots,
\\
{\bar C}_{0} &=&
-{\bar c}_{0}
- \lambda_i\ln\lambda_i
-(B-1)\lambda_i
+\ldots,
\\
\dz {\bar C}_{11} &=& \dz {\bar C}_{12} 
=
-({\bar c}_0 -B +1)(\lambda_i-1)+\ldots,
\\
\dz^2 {\bar C}_{23} &=&
-2 ({\bar c}_0 -B +1) + \frac{\dz}{2}
-~\left[\dz{\bar c}_0-4({\bar c}_0 -B +1)   \right]\lambda_i
+\ldots,
\\
2\dz {\bar C}_{24} &=&
- \frac{\dz}{D-4}
- ({\bar c}_0 -B +1) + \frac{3 \dz}{2} - \pi \dz y + 2\dz y\arctan(2y)
\nl && 
+ \left[2({\bar c}_0 -B +1)-\dz{\bar c}_0\right]\lambda_i
+\ldots,
\ea
with
\ba
B &=& 2y\left[\arctan(2y) + \arctan\left(\frac{\dz-1}{3-\dz}2y \right)
\right] 
\approx 1.75,
\label{B}
\ea
for $\dz = 1.286$,  $y \approx 0.73 $,  
and the values of $-C_0$ and $-{\bar C}_0$ at $\lambda_i = 0$
\cite{Mann:1984dvt}: 
\ba
\label{core}
\dz~c_0 &=& \frac{\pi^2}{6} - \litwo\left(\frac{1}{1+\dz}\right)
-\frac{1}{2}\ln^2(1+\dz) + \pi \ln(1+\dz) \times  i,
\\
\label{barcore}
\dz~{\bar c}_0 &=& \frac{\pi^2}{6} - \litwo(1-\dz) 
% next line a misprint in Annalen d. Physik. 19-11-99
+2 \Re e \litwo\left[(\dz-1)\left(\frac{\dz}{2}-1+\dz y \times i\right)
\right]
\nl &&
-~   2 \Re e \litwo\left(1-\frac{\dz}{2}-\dz y \times i\right)
\ea
Only the functions ${\bar C}$ develop imaginary parts, and only for 
\ba
\dz > 4\lambda_i.
\ea
The functions $C_0, C_{11}, {\bar C}_0$ contain terms of the
order $\dz\ln\dz$ at $\lambda_i \ll \dz$, but these terms cancel in the 
form factor $V$.
They survive in the case $\dz \to 0$ at constant $\lambda_i$.

A further note: the Euler dilogarithm is badly converging on the unit
circle. In fact, for one of the $\litwo$ above it occurs $|1-\dz/2 \pm i 
\times \dz y| = 1$. In that case, one may use:
\ba
\Re e \litwo\left( e^{i\phi}\right) &=& - \frac{1}{2} \litwo(1) +
\frac{1}{4} (\phi \pm \pi)^2, 
\ea
taking the value of $\phi$ fulfilling the condition:
\ba
(\phi \pm \pi) \in (-\pi,\pi).
\ea

The resulting small mass limit of the vertex is given in
\cite{Mann:1982xw,Riemann:1982rq} for the case of sequential Dirac neutrinos:
\ba
\label{small-vdi}
V(\lambda_i \ll 1, \lambda_Z) &=&
I_3^{i_L} \frac{8c_W^2}{D-4}
%+ V'(\lambda_i \ll 1, \lambda_Z),
%\\
%V'(\lambda_i \ll 1, \lambda_Z) &=& 
+
a_0 + a_L ~(\lambda_i \ln \lambda_i) + a_1 \lambda_i +
\ldots
\ea
The divergent constant was left out in the introduction,
Eqn. (\ref{smallD}).  

We want to stress that in this limit,
\ba
\label{small-vdL}
a_L = 0.
\ea
The ansatz in Eqn. (1) of
\cite{Clements:1983mk} allowing for $a_L\neq 0$ is too general for the
case of small neutrino masses in this respect.
Such a term occurs for $\lambda_Q = 0$, see section \ref{mapra}.

A numerical value of  $a_0$ was given in \cite{Mann:1982xw}.
The analytical expression is:
%%%%% general expression for $a_0$ is (unpublished so far, see notes
%   N82p.71,16-2-82-3 and corrected p.73):
 \ba
\label{a0ana}
a_0 &=&
\frac{(1+\dz)^2}{\dz}c_0 +\frac{2}{\dz^2}(1+2\dz)({\bar c}_0-B)
% term has wrong factor on page 71: +\frac{2}{3}[\pi y - 2y \arctan(2y)] 
+\frac{6}{\dz}[\pi y - 2y \arctan(2y)] 
\nl && 
+~\frac{2+3\dz}{2\dz}(\ln
\dz-\pi \times i) - \frac{1}{4\dz^2}(7\dz^2+14\dz-8).
\ea
The general expression for $a_1$ is \cite{Riemann:1982rq}:
\ba
\label{a1ana}
a_1(\dz) &=&
-\frac{2}{\dz}(1+\dz) c_0 
+ \frac{1}{2\dz^2}(4\dz^2-5\dz-6){\bar c}_0-\frac{2}{\dz}
( \ln \dz -\pi \times i)
\nl &&
+~\frac{1}{8\dz^2} (25\dz^2-38\dz-24) +\frac{1}{2\dz}(2-\dz)\pi y  
\nl && +~
\frac{1}{\dz^2}(\dz^2+7\dz+6)y\arctan (2y)  
\nl &&+~ \frac{3}{\dz^2} (3\dz+2) 
y\arctan\left(\frac{\dz-1}{3-\dz}2y\right),
\ea
with $y$ and $B$ from (\ref{y}) and (\ref{B}).
Further, we used $c_0$ and ${\bar c}_0$ of (\ref{core}) and
(\ref{barcore}).
With the inputs $M_W = 80.410$ GeV and $M_Z = 91.187$ GeV, these formulae
yield:
\ba
\label{a0}
% from 8-1-82-2, N82:
a_0 &=& 1.2584 + 1.0524\times i,
\\
\label{a1}
       a_1 &=&  2.5623  - 2.2950  \times i.
\ea
%         Old inputs:
% 
%         Re(Z) =  2.53337240709608 
%         Im(Z) = -2.31057549459323
% 
%In \cite{Mann:1982xw}, $a_0$ was given explicitly for $M_Z/M_W=1.25$
%and correspondingly $s_W^2=0.2$:
%\ba
%a_1(\lambda_Z=1.25) &=& 2.5334 -  2.3106 \times i.
%\ea
%=======================================================================
\subsection{\label{mapra}The vertex for $\lambda_Q=0$}
%=======================================================================
The first calculation of the non-diagonal $Zf_1f_2$ vertex seems to be
in \cite{Ma:1980px}, where the approach was 
simplified considerably by the approximation $Q^2=0$.
We should mention that this limit is of physical relevance 
only in the large neutrino mass limit, since only then it is
$\lambda_i\gg \dz \approx \lambda_Q=0$.
If we are interested in applications where $\lambda_Q> 4\lambda_i$,
%i.e. of order unity.
%If now the neutrino masses are very small, $\lambda_i<<1$, then they
%fulfill at the same time   
%\ba
%\lambda_i < \lambda_Z.
%\ea
%For this case, 
the amplitude gets essentially complex valued.
The limit of \cite{Ma:1980px}, however, implies automatically the relation 
$\lambda_i>\lambda_Q$, even for $\lambda_i \ll 1$, and the
amplitude is essentially real.
So one cannot expect a continuous behaviour for the small or 
medium neutrino mass range. The virtual fermions are considered Dirac
particles $(\lambda_i=\lambda_j)$.

The vertex 
%\footnote{In Eqn. (8) of \cite{Ma:1980px}, a different
%overall sign convention is chosen and the divergent part is suppressed.}: 
may be easily derived from the formulae given in
\cite{Mann:1984dvt}: 
% TR mentions the case first in \cite{Riemann:1982sx}
\ba
\label{mapratr}
{V}(\lambda_i,\lambda_Q =0) &=& I^{i_L}_3\left[
\lambda_i \left(\lambda_i-10\right) {\cal I}_1 +8{\cal L} +6
+ \frac{8c_W^2}{D-4} \right],
\ea
with 
\ba
\label{fun-i1}
{\cal I}_1 &=& \int_0^1 dx \frac{x}{(1-\lambda_i)x+\lambda_i}
= %\nl &=&
\frac{\lambda_i\ln\lambda_i}{(1-\lambda_i)^2}+\frac{1}{1-\lambda_i},
\\
\label{fun-l}
{\cal L} &=&\int_0^1 dx x\ln\left[(1-\lambda_i)x+\lambda_i \right]
= %\nl &=&
\frac{\lambda_i^2\ln\lambda_i}{2(1-\lambda_i)^2}+\frac{1}{2(1-\lambda_i)}
-\frac{1}{4}. 
\ea
The divergent part is independent of $\lambda_i$ and of $\lambda_Q$.

Explicitly:
\ba
\label{mapramudita}
{V}(\lambda_i,\lambda_Q =0) &=&
I^{i_L}_3\left(
3\frac{\lambda_i^2\ln\lambda_i}{(1-\lambda_i)^2}
+
2\frac{\lambda_i\ln\lambda_i}{(1-\lambda_i)^2}
-\frac{\lambda_i^2}{1-\lambda_i}
+6\frac{\lambda_i}{1-\lambda_i}
+\frac{8c_W^2}{D-4}
\right),
\nl
\ea
with $I^{i_L}_3$ being the weak isospin of the virtual
neutrinos.
The constant finite terms of the vertex vanish for small $\lambda_i$:
\ba
\label{mapra0}
{V}(\lambda_i \ll 1,\lambda_Q =0) &=& I^{i_L}_3\left(2\lambda_i\ln\lambda_i +
6\lambda_i+\frac{8c_W^2}{D-4}\right). 
\ea
This has to be compared to (\ref{small-vdi}) and  (\ref{small-vdL}).

The large mass limit is:
\ba
\label{maprabig}
{V}(\lambda_i \gg 1,\lambda_Q =0) &=& I^{i_L}_3\left(
 \lambda_i + 3\ln\lambda_i -5+\frac{8c_W^2}{D-4}
\right). 
\ea
This has to be compared to (\ref{vbig}) and  (\ref{bigva0}).

Finally, for the sake of completeness, the value of the vertex at the
weak scale: 
\ba
\label{mapra1}
{V}(\lambda_i =1,\lambda_Q =0) &=& I^{i_L}_3\left(\frac{3}{2}+\frac{8c_W^2}{D-4}
\right). 
\ea
\vspace{-5mm}
%=======================================================================
\def\href#1#2{#2}

%=======================================================================
\end{document}